\documentstyle[eqsecnum,aps,floats,epsfig,tabularx]{revtex}
\newcommand{\cd}{\cdot}

\newcommand{\al}{\alpha}

\newcommand{\de}{\delta}
\newcommand{\De}{\Delta}

\newcommand{\ga}{\gamma}
\newcommand{\Ga}{\Gamma}

\newcommand{\La}{\Lambda}
\newcommand{\la}{\lambda}
\newcommand{\Om}{\Omega}
\newcommand{\om}{\omega}
\newcommand{\si}{\sigma}
\newcommand{\Si}{\Sigma}
\newcommand{\th}{\theta}

\newcommand{\DD}{\mbox{$\cal D$}}
\newcommand{\GG}{\mbox{$\cal G$}}
\newcommand{\SS}{\mbox{$\cal S$}}

\newcommand{\Eq}[1]{ Eq.~(\ref{#1})}
\newcommand{\be}{\begin{equation}}
\newcommand{\ee}{\end{equation}}

\newcommand{\bea}{\begin{eqnarray}}
\newcommand{\eea}{\end{eqnarray}}
\newcommand{\bean}{\begin{eqnarray*}}
\newcommand{\eean}{\end{eqnarray*}}

\newcommand{\bk}{{\bf k}}
\newcommand{\bx}{{\bf x}}
\newcommand{\bp}{{\bf p}}

\newcommand{\bkp}{{\bf k'}}

\newcommand{\bbe}{{\bf e}}
\newcommand{\bw}{{\bf w}}
\newcommand{\bv}{{\bf v}}
\newcommand{\ie}{{\em i.e.~}}
\def\spose#1{\hbox to 0pt{#1\hss}} 
\def\ltapprox{\mathrel{\spose{\lower 3pt\hbox{$\mathchar"218$}} 
 \raise 2.0pt\hbox{$\mathchar"13C$}}} 
\def\gtapprox{\mathrel{\spose{\lower 3pt\hbox{$\mathchar"218$}} 
 \raise 2.0pt\hbox{$\mathchar"13E$}}} 
\def\inapprox{\mathrel{\spose{\lower 3pt\hbox{$\mathchar"218$}} 
 \raise 2.0pt\hbox{$\mathchar"232$}}}

\begin{document}
\draft
%\begin{flushright}
%CERN-TH/99-xxx \\
%UGVA-DPT 1999/04-zzzz\\
%astro-ph/0008232 \\
%\end{flushright}
%\vspace{15mm}
\preprint{\ 
\begin{tabular}{rr}
%CfPA/96-th-15 &  \\ 
& 
\end{tabular}
} 
%\twocolumn[\hsize\textwidth\columnwidth\hsize\csname@twocolumnfalse\endcsname 
\title{CMB anisotropies from pre-big bang cosmology}

\author{ F. Vernizzi$^1$, A. Melchiorri$^{3,2,1}$ and R. Durrer$^1$ }
\address{ $^1$D\'epartement de Physique Th\'eorique, Universit\'e de Gen\`eve,
24 quai E. Ansermet, 1211 Gen\`eve 4, Switzerland\\
$^2$ Universit\`a Tor Vergata, Roma, Italy\\
$^3$ NAPL, University of Oxford, Keble Road, Oxford, OX1 3RH, England}
\maketitle

\begin{abstract}
We present an alternative scenario for cosmic structure formation where
initial fluctuations are due to Kalb-Ramond axions produced during a 
pre-big bang phase of inflation. We investigate whether this scenario,
where the fluctuations are induced by seeds and therefore are of
isocurvature nature, can be brought in agreement with present
observations by a suitable choice of cosmological parameters. 
We also discuss several observational
signatures which can distinguish axion seeds from standard inflationary
models. We finally discuss the gravitational wave background 
induced in this model and we show that 
it may be well within the range of future observations. 

\end{abstract}
\date{\today}
\pacs{PACS Numbers : 98.80.Cq 98.80.E}

%%%%%%%%%%%%%%%%%%%%%%%%%%%%%%%%%%%%%%%%%%%%%%%%%%%%%%%%%%%%%%

\begin{section}{Introduction}

It is commonly assumed that an inflationary phase is necessary in 
order to construct a consistent cosmological model. 
The familiar adiabatic inflationary scenario 
owes its popularity to the fact that it 
solves the horizon and flatness problem
and at the same time provides a consistent model 
for the origin of cosmological perturbations.
In particular, it naturally leads to a flat
({\em Harrison-Zel'dovich})
spectrum of perturbations on large scale and to coherent acoustic
oscillations
on intermediate scales which manifest themselves as ``peaks'' in the 
Cosmic Microwave Background (CMB) anisotropies. 

After the recent measurements of the intermediate scale CMB anisotropy 
power spectrum \cite{knpa,toco,b97,debbe,max},  
flat adiabatic models seem to be favored \cite{dlkn,melk2k,teza}.
Nevertheless none of the many inflationary scenarios which have been 
developed during the last 20 years has 
been constructed consistently on the bases of a serious theory of
high energy physics; inflation has always been seen as an effective model
pointing to a greater more fundamental theory which has not been 
clarified so far. 
We  believe that superstrings are presently the most promising 
candidate for such a theory but on the other hand it is well known
that it is not possible to derive an inflationary model from a string theory 
effective action 
on a generic background, the reason being that the non-minimal 
coupling between the 
dilaton and the metric slows down the expansion of the universe 
spoiling the solution of the problems for which inflation 
has been invoked.

The pre-big bang idea \cite{PBB} represents in this context 
one of the first and most interesting
attempts to develop a new cosmological scenario which solves 
the horizon and flatness problems, based on string theory.
In this radically new picture, the underlying duality 
symmetry \cite{Brand} present in the low energy sector of string theory
naturally selects perturbative initial conditions and automatically leads 
to an inflationary phase prior to the big bang during which curvature
and the dilaton are growing \cite{PBB,G}.
Besides its many appealing features, this scenario 
is known to face several problems such as the
lack of a complete and consistent description of the high coupling and 
high curvature regime where the transition between the  
pre-big bang and the 
post-big bang phase and the stabilization of the dilaton should take place 
\cite{Exit}.
Furthermore, opinions vary as to whether the initial conditions in the 
pre-big bang need a 
large amount of fine tuning \cite{G,Tuning}. 
On a more phenomenological side, 
it is nevertheless important to study whether this scenario can 
provide the features 
that we observe in the universe today.
For recent review articles discussing several of the previous points
we refer the reader 
to \cite{Co,Ve}. For a comparison of the pre-big bang model with new 
cosmological models based on string theory see \cite{Ea}.

A realistic cosmological model has 
to generate large-scale matter perturbations and to reproduce the slope
and the amplitude of CMB anisotropy spectrum.
The pre-big bang scenario was thought for some time to be unable 
to provide a scale-invariant spectrum of 
perturbations. First-order tensor and scalar 
perturbations in the metric, as well as perturbations of the 
moduli fields,
were found to be characterized by extremely ``blue'' spectra \cite{5b}. 
This large tilt, together with a natural normalization 
imposed by the string cutoff at the shortest amplified scales, 
make their contribution to large-scale structure completely negligible.

However, it was later realized that the spectral tilt of the axion,
a universal field in string theory,
can assume a whole range of values depending on the behavior 
of the internal and external dimensions and in particular it 
can naturally provide a scale-invariant spectrum of perturbations 
\cite{Copeland,Buon,Hadad}.
This result reopened the possibility that pre-big bang cosmology 
may contain a natural mechanism for generating large-scale 
CMB anisotropies via the ``seed'' mechanism \cite{d90}. 

This possibility was analyzed in Refs. \cite{2,1} for massless axions and in
Ref.~\cite{massive} for very light axions. These analytical treatments are
restricted to large angular scales. 
We have extended  the study to smaller scales 
 with the help of numerical calculations. First results for this work
have been reported in a letter
\cite{afrg}, where a strong correlation between the axion spectrum, 
$n_{\si}$,
and the height of the peak was noticed.  
A range of values  around $n_{\si} = 1.4$ (slightly blue spectra) 
appeared to be favored  by a simultaneous fit to
the normalization on large angular scales observed by
COBE \cite{COBE} and the data on the first acoustic peak 
available at that time.

In this companion paper we present a full explanation
of the  details of these calculations for  
the CMB angular power spectrum and for the dark matter power spectrum  
and we study the problem of the ``decoherence'' 
of axion perturbations which has been ignored in the previous work.
Furthermore, we expand on the results on the observational signatures
presented  in the Letter~\cite{afrg} and we discuss them in the light of
the new CMB anisotropy data presently available 
by  investigating the cosmological
parameter-space of the model. We also 
discuss CMB polarization for our model and the contribution of the
gravitational wave background induced by axion perturbations.

We will suppose for the rest of this work that  
modes with frequencies relevant to CMB physics
are unaffected by the 
the transition from the pre- to the post-big bang phase.
This may seem a strong assumption given our ignorance on 
the graceful exit and on the duration of the intermediate string phase.
Indeed, the nature and features of the transition 
are still unknown and may lead to important changes on the 
characteristics of the axions spectrum, but we are confident 
that this can only affect scales much smaller than those  
we are interested in: For this to hold, we just need 
that, at the moment we enter the string phase,
the CMB modes are well outside the horizon and therefore not altered
by any causal process. 
As shown in~\cite{Dua} using general arguments, the long-wavelength
part of the solution to the 
equation of motion of the axion field is always dominated 
by the contribution of the frozen modes even if the background evolution 
includes a high-curvature phase in which the action 
and the perturbation equations are not known.
This argument has been tested by numerical examples in~\cite{Gasp} 
for a gravitational wave background spectrum.
Recently a graceful exit model considering general 
high-curvature and coupling corrections has appeared in the literature
\cite{Cyril} and numerical 
tests on the axion spectrum have been performed and will be published
soon~\cite{CF} to confirm our assumption.

The paper is organized as follows: in the next section we study axion
production in the pre-big bang and explain the details of the
computation of the axion energy-momentum tensor which plays the role
of the ``seed'' in our model. In Section~III we determine the CMB
anisotropy and dark matter spectra. We study the problem of
decoherence and show that the coherent approximation is very good for
this model. In Section~IV we compare our result with CMB and Supernova data and
present a cosmological parameter estimation for this scenario. We also
examine and discuss the normalization and the kink
in the axion spectrum which is required to fit
observations. Section~V is devoted to a novel prediction of axion
seeds: the tensor component of their energy-momentum tensor induces a
gravity wave background which might be observable.
In Section~VI we summarize our conclusions.

\end{section}
\vspace{1cm}
%%%%%%%%%%%%%%%%%%%%%%%%%%%%%%%%%%%%%%%%%%%%%%%%%%%%%%%%%%%%%%

\begin{section}{Axion seeds from string cosmology}

\begin{subsection}{Extra dimensions in string cosmology}

The minimal low energy effective 
action of the NS-NS sector of string theory in the string frame is
given by~\cite{String}
\be
S_{10} = \int d^{10}x \sqrt{|g_{10}|} e^{-\phi_{10}} \left[R_{10} +
 (\nabla \phi_{10})^2 -{1 \over 12}H_{10}^2 \right],
 \label{Saction}
\ee
where we have included the 10-dimensional antisymmetric tensor
$H_{\mu\nu\alpha}=\partial_{ [\mu} B_{ \nu \al] }$, but no gauge or
fermion fields.

We assume that the 10-dimensional metric can be factorized into a
``large'' four-dimensional part and a ``small'' six-dimensional metric,
\be
ds^2=g_{\mu \nu} dx^{\mu} dx^{\nu} 
+ e^{2 \beta}\delta_{IJ} dX^I dX^J
\ee
($\mu,\nu=0,\ldots,3$ and $I,J=1,\ldots,6$),
where $\beta$ depends only on time, $\beta=\beta(t)$. 
If the six dimensional
piece is compactified to a very small radius, the lowest energy
Kaluza-Klein modes yield  the four-dimensional action \cite{Copeland},
\be
S = \int d^{4}x \sqrt{|g|} e^{-\phi} \left[R +
 (\nabla \phi)^2 -3(\nabla \beta)^2 
-{1 \over 2}e^{2\phi}(\nabla \si)^2 \right].
 \label{S4action}
\label{Action4}
\ee
Here we have introduced the four-dimensional 
 axion field $\sigma$ defined by 
\be
H^{\mu\nu\alpha} =
e^{\phi} \epsilon^{\mu\nu\alpha\beta} \nabla_{\beta} \sigma.
\label{Baxduality}
\ee
The action~(\ref{S4action}) and the definition~(\ref{Baxduality}) 
include the four-dimensional dilaton field, $\phi$, 
the pseudo-scalar axion field, $\si$, which represents the degrees of
freedom of the antisymmetric three tensor field $H$, and a modulus field, 
$\beta$, which parameterizes the radius, 
or the ``breathing mode'', of the six-dimensional internal space.   
Like the dilaton, also 
the axion field (not to be confused with the Peccei-Quinn axion) 
is universal in string theory.

Let us assume a homogeneous dilaton background, $\phi=\phi(t)$,  
and an external four-dimensional spacetime 
adequately described by a standard, spatially flat FLRW metric 
with scale factor $a(t)$,
\be
g_{\mu \nu}=\mbox{diag}[-1,a^2(t),a^2(t),a^2(t)].
\ee
In the following we shall also make use of the metric 
\be
g_{\mu \nu}=a^2(\eta)\mbox{diag}[-1,1,1,1],
\ee
where we have introduced the conformal time $\eta$ given by
$d\eta=dt /a$ (we shall use an over-dot to indicate a derivative with
respect to conformal time, $\dot{} \equiv \partial/\partial \eta$).
With our choice of the external metric, the four-dimensional dilaton 
is related to the 10-dimensional one by
\be
\phi=\phi_{10} - 6\beta.
\ee

When the axion field is trivial, $\dot{\sigma}=0$, or its 
contribution to the global dynamics of the universe is negligible, 
the equations derived from the action (\ref{Action4})
are invariant under 
duality transformations,
\be
a(t) \rightarrow 1/a(-t), \ \ \ \ \ \phi(t) \rightarrow 
\phi(-t) - 6\ln(a(-t)).
\ee
This invariance ({\em scale factor duality}) represents one of the
key motivations behind the pre-big bang scenario~\cite{PBB}.
The field equations for $a,\phi$ and $\beta$ are solved~\cite{PBB}
by the following power laws, known as {\em dilaton-vacuum} solutions in
the pre-big bang for $\eta<-\eta_1$: 
\be
a(\eta)= \left({-\eta\over\eta_1}\right)^{\frac{\delta}{1-\delta}}, \ \ \ \ 
e^{\beta(\eta)} =  
 \left({-\eta\over\eta_1}\right)^{\frac{\zeta}{1-\delta}}, \ \ \ \ 
e^{\phi(\eta)} = 
	\left({- \eta\over\eta_1}\right)^{\frac{3 \delta -1}{1-\delta}},
\label{Solu}
\ee
where $\delta$ and $\zeta$ satisfy  the Kasner constraint,
\be
3 \delta^2 + 6 \zeta^2 =1.
\label{Kasner}
\ee
Here $-\eta_1$ is the (conformal) time at which curvature and 
dilaton become so large that loop corrections from string theory have 
to be taken into account. It is hoped that these corrections then lead
to a radiation dominated Friedman universe with ``frozen'' dilaton
at $\eta > \eta_1$.
From these solutions one can see that, during the pre-big bang phase, \ie
for negative conformal time $\eta$, 
a negative $\delta$ and a positive $\zeta$ are required 
to make the external three-dimensional space expand and the internal six-dimensional
space contract. Therefore $\delta$ has to lie in the interval 
 $-1/\sqrt{3} \le \delta < 0$,
which leads  always to a growing dilaton and growing four-curvature, $R
\sim (\dot a/ a^2)^2 \propto 1/(a\eta)^2\propto  (-\eta)^{-2\over 1-\de}$.

\end{subsection}

\begin{subsection}{Amplification of axion quantum fluctuations}
\label{Ampli}
In this subsection we briefly review the mechanism 
for the generation of a primordial quasi-scale-invariant spectrum
from the pre-big bang phase and we discuss 
the dependence of the spectral index on the
evolution of the internal and external dimensions of the 
pre-big bang universe.
Using as initial conditions the axion field obtained during
the  pre-big bang phase, we then analyze its evolution after 
the big bang in a critical FLRW 
universe with and without cosmological constant, paying
particular attention to the frequency modes that enter 
into the calculation of the CMB anisotropy power spectrum.

As in  
previous works \cite{2,1,massive,afrg} we suppose that the contribution 
of the axion field to the equations of motion
for $\phi$, $a$ and $\beta$ is negligible and that
the evolution of the dilaton, the moduli, and the scale  factor are
governed by the dilaton-vacuum solutions (\ref{Solu}).   
Nevertheless, quantum fluctuations of all the fields are of course
present and we will show that quantum fluctuations of the axion field
can  seed density perturbations and  CMB anisotropies in the post-big
bang era. To this goal we have to study the axion evolution equation and the
spectrum of axions produced during the pre-big bang phase due to their
coupling to the background gravitational field and the dilaton.

Varying the action (\ref{Action4}) with respect to the field $\sigma$
in the string frame yields the equation of motion
\be
\nabla_\mu (e^{\phi} \nabla^\mu \sigma)=0.
\ee 
The study of this equation is conveniently performed by using the
{\em canonical variable} given by
\be
\psi\equiv a_A \sigma \equiv ae^{\phi/2}\sigma, 
\ee
which ``diagonalizes'' the perturbed action 
expanded up to second order. The factor 
$a_A$ is the so called {\em pump field} of the axion.
The Fourier
modes $\psi_{\bk}(\eta)$ satisfy a canonical linear second-order
equation, completely decoupled from the other fields,
\be
 \ddot{\psi_{\bk}} +
\left(k^2-{\ddot{a}_A\over a_{A}}\right)\psi_{\bk} =0 . 
\label{Evol}
\ee 
This is the {\em evolution equation} for the axion field.

\Eq{Evol} is equivalent to the equation for a classical harmonic oscillator 
with parametric evolution driven by the time
dependent {\em mass term}
$\ddot{a}_A/a_A$. 
When the time evolution of the velocity of the pump field, $\dot{a}_A$,
is sufficiently slow such that, for a given mode $k$, 
$\ddot{a}_A/a_A \ll k^2$,
we are in the adiabatic regime with the result that no particles are created.
When the acceleration in the pump field is 
high enough to violate the adiabatic regime, quantum particle 
production starts.
The evolution of the axion field and the resulting
spectrum of particles are fully determined by the time behavior of the pump 
field in the different phases of the universe. 
In particular, a strong difference in this behavior exists between 
the pre-big bang phase and the standard radiation and matter dominated eras
in the post-big bang universe.

The pre-big bang phase is 
characterized by an accelerated evolution of the pump field, 
\be
a_A \propto (-\eta)^{\gamma},
\ \ \ \ \ 
\gamma=\frac{5\delta-1}{2(1-\delta)},
\label{Gammadelta}
\ee
where $\delta<0$ is the power 
which characterizes 
the evolution of the external dimensions, \Eq{Solu}.
Using \Eq{Evol}, the evolution equation of the axion can be
written as 
\be
\ddot{\psi}_{\bk}+k^2 \left(1-\frac{\gamma(\gamma-1)}{x^2} 
\right)\psi_{\bk}=0, 
\label{Evolpre}
\ee
where $x \equiv k\eta$.
This equation is solved analytically   in terms of the 
Hankel functions $\eta^{1/2}H^{(1)}_\mu$ and $\eta^{1/2}H^{(2)}_\mu$
with $\mu=|\ga-1/2|$. 

At very early time,  a perturbation of given wave number $k$ is well
inside the horizon, $|x| = |k \eta| \gg 1$, and the solutions of
\Eq{Evolpre} are harmonic oscillations which 
can be consistently normalized to the vacuum fluctuation spectrum for 
$\eta \to - \infty$. This initial condition implies that the
$H^{(1)}_\mu$ mode is absent and
\be
\psi_{\bk}(\eta)=(-\eta)^{1/2} H^{(2)}_{\mu} (k \eta), \ \ \ \ \ 
\mu= \frac{1}{2} - \gamma = \frac{1-3 \delta}{1- \delta}, 
\ \ \ \ \ \mbox{for} \ \ \eta < -\eta_1.
\label{Solpre}
\ee
At $\eta = -\eta_1$ we ``glue'' the pre-big bang phase to a radiation dominated post-big bang era starting at $\eta=\eta_1$. 

After the singularity, 
during the standard radiation and matter dominated eras, the dilaton 
is frozen, $\phi=\mbox{const}$, and the pump field is proportional 
to the standard
scale factor, $a_A \propto a$.
The scale factor, $a$, and its second derivative, $\ddot{a}$, are
given by Friedman's equations. For a critical universe, which we
consider throughout our calculations and which is certainly a good
approximation until redshifts $z\le 5$, we have 
\bea
\frac{\ddot{a}}{a}&=&\frac{4\pi G}{3}  a^2 (\rho - 3p)
+\frac{2a^2\Lambda}{3}, 
\label{Ein1} \\
\frac{\dot{a}^2}{a^2} &=&\frac{8\pi G}{3} a^2 \rho
+\frac{a^2\Lambda}{3}~.
\label{Ein2}
\eea 
Energy conservation for radiation ($_r$) and
 matter ($_m$) yields $\rho_{r} \propto 1/a^4$
and $\rho_{m} \propto 1/a^3$, with
$\rho = \rho_{r} + \rho_{m}$ and $p=\rho_{r}/3$; $\rho_r$ is the 
radiation energy-density, $\rho_m$ is the matter energy-density, and $p$
the pressure of the radiation fluid.
At early times, when $\Lambda$ is negligible, these equations have a
simple analytical solution,
\be
a = a_{eq}\left( \eta/\eta_* + \frac{1}{4}(\eta/\eta_*)^2\right),
\ \ \ \ \ \eta_* \equiv \left( \frac{3}{4 \pi G \rho_{eq}} \right)^{1/2}
 =  \frac{\eta_{eq}}{2(\sqrt{2}-1)} \simeq 1.2 \eta_{eq},
\label{Effpot}
\ee
where $\eta_{eq}$ is the transition time between the  radiation
and the matter dominated era, $\rho_r(\eta_{eq}) =
\rho_m(\eta_{eq})= \rho_{eq}/2$.
The mass term during the post-big bang becomes

\be
\frac{\ddot{a}_A}{a_A}=
\frac{\ddot{a}}{a}={1\over 2\eta \eta_* + {1\over 2} \eta^2 }.
\label{Effpotan}
\ee
When $\Lambda$ is non vanishing, the solution for the 
mass term can be found numerically but since the 
contribution of a small cosmological
constant to the scale factor
becomes important only at late time, the solutions to  (\ref{Evolpre})
are almost unaffected; this has been checked by numerical tests. 
In the radiation dominated era, $\eta < \eta_{eq}$, 
the mass term can be approximated by 
${\ddot a}_A/a_A \simeq 1/(2\eta_*\eta)$.

We now study the axion evolution in the post-big bang era.
Let us write the term in parenthesis on the left hand side of  
the axion equation of motion, \Eq{Evol}, as
\be 
\left(k^2 -\frac{\ddot{a}}{a} \right) = 
k^2 \left(1-\frac{(\ddot{a}/a)\eta^2}{x^2} \right) =
k^2 \left(1-\frac{\eta/(2\eta_*+\eta/2)}{x^2}  \right).
\ee
In order to study the solution of \Eq{Evol} we have to study the ratio of
the {\em dimensionless mass term} $(\ddot{a}/a)\eta^2$ and $x^2$
to be compared with unity.
As long as we are well in the radiation dominated era, $\eta \ll \eta_*$, the 
dimensionless mass term  is small and particle
creation induced by the pump field is negligible at early times.
 \Eq{Evol} then is a harmonic equation solved by free plane waves,
\be
\psi_{\bk}(\eta)=\frac{1}{\sqrt{k}}[c_+(\bk)e^{-ik\eta}+c_-(\bk) e^{ik\eta}].
\label{Solrad} 
\ee 
By matching the two solutions (\ref{Solpre}) and (\ref{Solrad}) 
at the transition time $\eta_1$
we obtain, for $|k \eta_1| \ll 1$ and $\eta_{eq} \gg \eta > \eta_1$, 
\be
	c_{\pm}(\bk)=\pm c(\bk), ~\mbox{ with }~~~
	\langle |c(\bk)|^2 \rangle \simeq \left\{\begin{array}{ll}
       \left( \frac{k}{k_1} \right)^{-2\mu-1} & k <k_1 \\
        0  & k>k_1~, \end{array} \right.  \label{bog}
\ee
so that
\be
\psi_{\bk}=\frac{c(\bk)}{\sqrt{k}} \sin (k(\eta - \eta_1)),
\ \ \ \ \ \mbox{for} \ \ \eta_1 < \eta \ll \eta_{eq}~.
\label{Presol}
\ee
Here $k_1=1/\eta_1$ represents the maximal amplified 
frequency of the pre-big bang phase. 
As already discussed in the introduction we suppose that  
modes with frequencies much lower than $k_1$ are unaffected by the 
unknown details of the transition from the pre- to the post-big bang phase.

The energy-density distribution of the produced axions is then
\be
\frac{d\rho_{\sigma}(\bk)}{d\log k} \simeq {1\over \pi^2}\left( \frac{k}{a} \right)^4
\langle |c(\bk)|^2 \rangle \simeq \left( \frac{k_1}{a} \right)^4 
\left( \frac{k}{k_1} \right)^{3-2\mu} \propto k^{n_{\sigma}-1}.
\ee 
 The axion spectral index $n_{\sigma}$ is related to the power 
which characterizes the evolution of the external dimensions by
\be 
n_{\sigma}=4-2\mu=3+2\gamma=2\left(\frac{1+\delta}{1-\delta}\right),
\label{Nsigma}
\ee
which follows from \Eq{Solpre}.
In order not to over-produce infrared axions we have to require 
$\mu \leq 3/2$, or $n_{\sigma} \geq 1$, which implies $\de\ge -1/3$.
As already pointed out in \cite{2},
the limiting value $\mu=3/2$ corresponds precisely to a 
Harrison-Zel'dovich spectrum of CMB anisotropies on large scale.
In terms of the evolution of the scale factor, this corresponds to
an isotropic expansion and contraction respectively of the external and 
internal dimensions,
\be 
a \propto \frac{1}{e^{\beta}} \propto (-\eta)^{-1/3}.
\ee
Notice that 
only for a 10-dimensional spacetime, 
symmetrical expansion and contraction corresponds to a
flat axion spectrum which induces a Harrison-Zel'dovich spectrum of
CMB fluctuations~\cite{1,2}! 

Nevertheless, as will be discussed in Section~IV, 
at very large scales and very early (negative) 
times, we will need a
slightly blue axion spectrum to fit CMB data. This requires a somewhat
larger value of $\de$, {\em i.e.} a slower expansion of the  external 
dimensions and, correspondingly, a
somewhat faster contraction of internal dimensions at  early time.
This blue spectrum cannot be maintained up to the string scale because the 
fixed normalization at the string scale to 
$g_1^2=[(k_1/a_1)/m_{\rm Planck}]^2 \sim 0.01 \div 10^{-4}$ would lead to much too
small amplitudes at the COBE scale.

Let us therefore investigate what happens if the universe expands with
some expansion law described by $\de_-$ at early times, $\eta<\eta_b
>-\eta_1$ and then switches to an expansion law given by $\de_+$ after
$\eta_b$. Sufficiently short wavelength
modes which are inside the horizon during the entire epoch
$\eta<\eta_b$, which satisfy $|k\eta_b| <1$, are not
influenced by this change in the expansion law. The term
$\ddot{a}_A/a_A$ is indeed sub-dominant in the equation of motion for 
$\psi_{\bk}$
during this epoch and hence the 
Bogoliubov coefficient $|c(\bk)|^2$ of Eq.~(\ref{bog}) is
not influenced by the transition;  we just obtain
the result~(\ref{bog}) with $\mu=\mu_+$. 

The situation is different if a mode exits the horizon before
$\eta_b$. Then the ``incoming'' solution
$\psi(\eta<\eta_b)=
(-\eta)^{1/2}H^{(2)}_{\mu_-}(k\eta)$ differs from the vacuum solution and
matching it to the general ``outgoing'' solution, 
$\psi(\eta>\eta_b)=b_1(-\eta)^{1/2}H^{(1)}_{\mu_+}(k\eta) + 
	b_2(-\eta)^{1/2}H^{(2)}_{\mu_+}(k\eta)$, 
yields $b_2-b_1 =
\left(\Ga(\mu_-)/\Ga(\mu_+)\right)|k\eta_b/2|^{\mu_+-\mu_-}$. 
Correspondingly, the coefficient $|c(\bk)|^2$ is changed by a factor 
$|b_2-b_1|^2$. In a model where the expansion law changes at a well
defined time $\eta_b\equiv -1/k_b$, we therefore get the following
Bogoliubov coefficients in the post-big bang radiation era 
(see Fig.~\ref{spettrof}):
\be
\langle |c(\bk)|^2 \rangle \simeq \left({k\over
                                        k_1}\right)^{-1-2 \mu_+}
\left\{
 \begin{array}{ll}
 (k/k_b)^{2\mu_{+} -2 \mu_-} & \mbox{for }~ k\le k_b \\
  1 & \mbox{for }~ k\ge k_b . 
 \end{array} \right.
\ee

\begin{figure}[ht]
\centerline{\epsfig{file=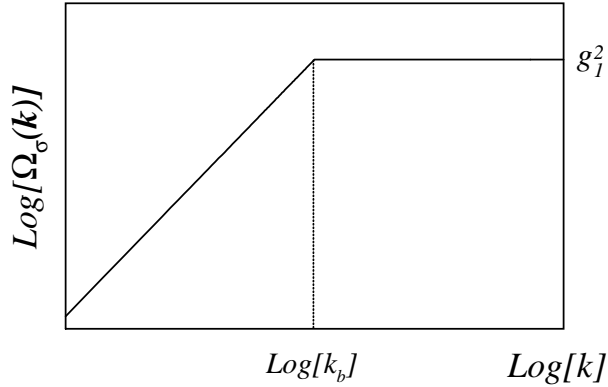, width=3.5in}}
\caption{Evolution of the axion spectral index  during the pre-big
bang. The value $g_1$ is the string coupling constant given by
the string scale divided by the Planck scale, $g_1=(k_1/a_1)/m_{\rm
Planck}$ (see Section~IV).
\label{spettrof}}
\end{figure}
We do not want to specify the event which may have triggered such a
transition from $n_{\sigma}(k < k_b)=4-2\mu_-=1+\varepsilon$ to 
$n_{\sigma}(k > k_b)=1$, 
but there are
certainly different possibilities. For example, 
it is interesting to note that
isotropic expansion and contraction, $a \propto 1/b$, 
in a $26$-dimensional spacetime
gives $\de_-=1/5$, or $n_{\sigma}=1.33$, 
which corresponds to $\varepsilon=1/3$, just about the
``tilt'' needed to fit the observed CMB anisotropies (see
Section~III). Therefore, if we start out the pre-big bang phase 
with a 26-dimensional bosonic
string vacuum (which we know to be unstable due to the presence of 
tachyons) which then ``decays''  to a supersymmetric and 10-dimensional
string vacuum at some
time $\eta_b$, which corresponds to a comoving energy scale $k_b$, this
could induce the required tilt. 

We now study the modification in the axion spectrum during the
post-big bang  era, where $a_A=a$. As we have seen above, during the
radiation  era, $\eta < \eta_*$, the 
dimensionless mass term  is small. Furthermore, once a mode
enters the horizon, $k\eta>1$, the $k^2$-term always dominates over
the mass term and there is no more particle creation. Therefore
modes which enter  the horizon before equality, $k \eta_* \gtapprox 1$, are
not amplified any further in the post-big bang phase.
The spectrum of axion perturbations for these modes remains unaffected.
However, the low frequency tail of the spectrum is further modified as soon 
as we enter in the matter dominated era, where the dimensionless
mass becomes of order unity. The 
 modes which enter the horizon after equality, 
$k \eta_* \ltapprox 1$, are amplified. This amplification of low
frequency  modes  has important consequences on the angular 
spectrum of the CMB as  will shall discuss in detail in Subsection~\ref{deco}.

\begin{figure}[ht]
\centerline{\epsfig{file=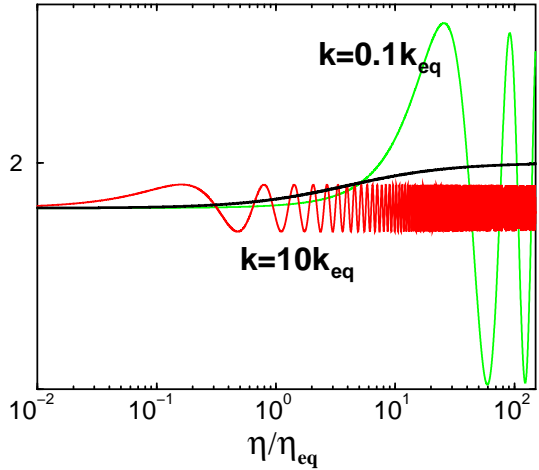, width=3.5in}}
\caption{The dimensionless mass term, $(\ddot{a}/a)\eta^2$, 
(thick line) and two modes that enter the horizon 
before and after equality.
The mode which enters the horizon before equality, 
$\varphi(k=10 k_{eq},\eta)$, is unaffected by the pump field 
and begins to oscillate without being amplified.
The mode that enters the horizon after equality, 
$\varphi(k=0.1 k_{eq},\eta)$, is amplified 
and begins to oscillate later.}
\label{potential}
\end{figure}

The behavior of the dimensionless mass term $(\ddot{a}/a)\eta^2$
together with two modes that enter the horizon 
before and after equality
have been plotted in Fig.~\ref{potential}.
As one can see, only modes entering the horizon after equality  
are amplified.  Deep in the matter era
$\eta \gg \eta_{eq}$, the dimensionless mass term is constant and 
\Eq{Evol} becomes
\be
\ddot{\psi}_{\bk}+\left(k^2 -\frac{2}{\eta^2}\right)\psi_{\bk}=0.
\ee  
This equation can again be solved in terms of Hankel functions,
\be
\psi_{\bk}(\eta)=\eta^{1/2} [AH_{3/2}^{(2)}(k\eta) +
BH_{3/2}^{(1)}(k\eta)], \ \ \ \  \mbox{for} \ \ \ \eta \gg \eta_{eq},
\label{Solmoto2}
\ee
where $A$ and $B$ are constants to be determined
by matching conditions (see \cite{1}). 
The post-big bang solutions (\ref{Presol}) 
and (\ref{Solmoto2}) are only correct far from 
matter-radiation equality $\eta_{eq}$ and in order to compute CMB anisotropies
we require better precision for these solutions also for 
$\eta \sim \eta_{eq}$.  We therefore solve the
axion equation of motion~\Eq{Evol} numerically, from the early
radiation era through the radiation-matter transition.

The axion field is then given by
\be
\sigma(\bk,\eta)=\frac{1}{a(\eta)}\psi_{\bk}(\eta)=
\frac{c(\bk)}{a\sqrt{k}} \varphi(k,\eta),
\ee
where the variable $\varphi$ is the 
solution of equation
\be
\ddot{\varphi}+\left(k^2 - \frac{\ddot{a}}{a} \right) \varphi=0
\label{Evol2}
\ee
with  initial condition (obtained from the pre-big 
bang solution \Eq{Presol})
\be
\varphi(k,\eta)=\sin (k\eta), \ \ \ \ \ \eta \ll \eta_*.    
\ee
We have solved \Eq{Evol2} numerically in this work
using the mass term (\ref{Effpotan}). The pre-factor 
$c(\bk)$ is a stochastic Gaussian field with power spectrum
\be
\langle |c(\bk)|^2 \rangle  = (k/k_1)^{n_{\sigma}-5},
\ee
where $n_{\sigma}$ is again the primordial spectral index (\ref{Nsigma}),
our free parameter which depends on the higher-dimensional 
pre-big bang phase.

\end{subsection}

%%%%%%%%%%%%%%%%%%%%%%%%%%%%%%%%%%%%%%%%%%%%%%%%%%%%%%%%%%%%%%%%%%%%%%%%%%%%%

\begin{subsection}{Axion quantum fluctuations as seeds}
\label{source}

We are now ready to consider the axion field as a source
of the linear cosmological perturbation equations.
As in previous works \cite{2,1,massive,afrg} we suppose that 
the contribution of the axions to the cosmic fluid can
be neglected and that they interact with it only
gravitationally. They then play the role of seeds which, by their
gravitational field, induce fluctuations in the cosmic fluid\cite{d90}.
The back-reaction of the metric perturbations on the evolution of 
seeds is second order and can be neglected in first order perturbation
theory. The evolution of axions can be computed 
by using the solutions of the axion field
equation in the unperturbed background geometry, \Eq{Evol}.

The axion field $\sigma$ is a Gaussian stochastic variable.
Its contribution 
to the perturbation equations is given in terms of 
its energy-momentum tensor,
\be
T_{\mu\nu}^{(\sigma)}=\partial_{\mu}\si\partial_{\nu}\si
-\frac{1}{2} g_{\mu \nu}
(\partial_{\al}\si)^2,
\label{Emt}
\ee
which is quadratic in $\sigma$ and 
therefore not Gaussian. Moreover, although the axion field evolves 
according to a linear equation,
it will enter into the perturbation equations 
through $T_{\mu\nu}^{(\sigma)}$ which evolves non-linearly.

The perturbations in the dark matter and radiation components are set 
to zero in the initial conditions and are subsequently 
induced by the gravitational field of the axion. Hence,
axion seed perturbations
belong to the class of {\em isocurvature perturbations}. However, they
differ from topological defects by being {\em ``acausal''}, {\em i.e.} 
they have non-vanishing correlations on super-Hubble scales, since they
are due to field excitations induced during an inflationary era. 

As we
have seen above, the axion power spectrum  obeys a simple power law with
cutoff and is
in general not analytic at $k=0$. Furthermore, axion perturbations do not,
in general, display the scaling behavior expected from topological
defects. In the pre-big bang we have an additional scale, the string
scale $k_1$, which breaks scale-invariance. The axion spectrum on large
scales is therefore not determined by dimensional arguments since
there are dimensionless factors of the form $(k/k_1)^{\alpha}$ which may
alter the spectrum\footnote{Actually the radiation -- matter
transition scale $\eta_*$ represents a scale which is also present in
models with topological defects, but deep in the radiation or matter
era this scale has no significance, whereas as
we shall see the above factors multiply the entire power spectrum of fluctuations.}.
The significance of these points will become clearer later in the paper.

As in \cite{afrg}, 
 we first consider a critical universe (total 
density parameter $\Omega=1$)
consisting of cold dark matter, baryons, photons, and three types
of massless neutrino, with or without a cosmological constant.
%($\Omega_{\Lambda}=0.7$ or $\Omega_{\Lambda}=0.0$). 
We choose the baryonic density parameter 
$\Omega_b=0.05$ and the value of the Hubble parameter 
$H_0=100h\mbox{km}\mbox{s}^{-1}\mbox{Mpc}^{-1}$ with $h=0.65$.

The linear perturbation equations for this universe 
in Fourier space are of the form
\be
 \DD X = \SS, \label{diff}
\ee
where $X$ is a long vector containing all the fluid perturbation
variables which depends on the wave number $\bk$ and conformal time
$\eta$, $\SS$ is a source vector which consists of certain
combinations of the seed energy momentum tensor and $\DD$ is a linear
ordinary differential operator.
More details on the linear system of
differential equations~(\ref{diff}) can be found in \cite{DKM}
and references therein.

For a given initial condition, this equation can 
in general be solved by means
of a Green's function, $\GG(\eta,\eta')$, in the form
\be
X(\bk,\eta_0) = \int_{\eta_{in}}^{\eta_0}\GG(\bk,\eta_{0},\eta)
	\SS(\bk,\eta)d\eta .
\label{Green}
\ee
We want to determine power spectra or, more generally, quadratic
expectation values of the form
\be
\langle X_i(\bk ,\eta_0)X_j(\bk,\eta_0)^*\rangle,
\ee
which, according to \Eq{Green}, are given by
\be
\langle X_i(\bk ,\eta_0)X_j(\bk,\eta_0)^*\rangle 
=\int_{\eta_{in}}^{\eta_0}\int_{\eta_{in}}^{\eta_0}
	\GG_{il}(\eta_{0},\eta)\GG_{jm}^*(\eta_{0},\eta') 
%\times \nonumber \\
	\langle \SS_l(\eta)\SS_m^*(\eta')\rangle d\eta d\eta' .
\label{pow}
\ee
(Sums over double indices are understood.)

We therefore have to compute the {\em unequal time correlators},
$\langle \SS_l(\eta)\SS_m^*(\eta')\rangle$, of the seed energy-momentum
tensor. This problem can, in general, be solved by an eigenvector
expansion method~\cite{DKM,Turok}, as it will be done in 
Subsection~\ref{deco}.
However, if the source evolution is linear, the problem becomes
especially simple. In this ``coherent'' case, we have
\be
 \SS_j(\eta) =F_{ji}(\eta,\eta_{in})\SS_i(\eta_{in}),
\label{Transfert}
\ee
with a deterministic transfer function $F_{ji}$. In this situation
we can, by a simple change of variables, diagonalize the hermitian,
positive initial equal time correlation  matrix,
\[
 \langle \SS_l(\eta_{in})\SS_m(\eta_{in})\rangle =\la_l\de_{lm}.
\]
Inserting this in \Eq{pow} yields
\be
\langle X_i(\eta)X_j^*(\eta')\rangle = \left(\int_{\eta_{in}}^{\eta_0}
   \GG_{il}(\eta_{0},\eta)F_{il}(\eta,\eta_{in})\sqrt{\la_l}d\eta \right)
 \left(\int_{\eta_{in}}^{\eta_0}
   \GG_{jm}(\eta_{0},\eta')F_{jm}(\eta',\eta_{in})\sqrt{\la_m}d\eta'\right)^* \delta_{lm}.
\label{coherent}
\ee
We therefore obtain exactly the same result as the one obtained by
replacing the stochastic variable $\SS$  by the deterministic 
source term $ \SS_j^{(det)}$ given by
\be
 \SS_j^{(det)}(\eta)\SS_i^{(det)}(\eta) =F_{jl}(\eta,\eta_{in})
	F_{il}(\eta,\eta_{in})\la_l = \exp(i\th_{ji})
 \sqrt{\langle |\SS_j(\eta)|^2\rangle\langle |\SS_i(\eta)|^2\rangle} ,
\label{Coerente}
\ee
where $\th_{ji}$ is a, in principle unknown, phase which has to be
determined case by case. Clearly $\th_{jj}=0$.
When the stochastic variable $\SS$ is real (as in our case) 
$\exp(i \theta_{ji})=\pm 1$.
This linear or coherent  approximation will be fully used in this paper.
We shall test its validity in Subsection~\ref{deco}.

It is useful to split the energy-momentum tensor of the axion seeds 
(\ref{Emt}) into a scalar, 
vector, and tensor part since the perturbations generated by each of
these components evolves independently. 
Due to statistical isotropy these three modes are uncorrelated.
This also corresponds to a decomposition of the source 
term $\SS$ into a scalar, 
vector,  and tensor contributions, $\SS^{(S)}$, $\SS^{(V)}$, and  $\SS^{(T)}$.
A suitable parameterization of the  decomposition of 
the Fourier components of $T_{\mu\nu}^{(\sigma)}$
is \cite{d90}
\bea
T_{00}^{(\sigma)}&=&f_{\rho}, \nonumber \\
T_{j0}^{(\sigma)}&=&-ik_j f_{v} +v_j, 
\label{Param} \\
T_{ij}^{(\sigma)}&=& \delta_{ij} f_p -
\left( k_i k_j - \frac{k^2}{3}\delta_{ij} \right)f_{\pi}
+ \frac{1}{2} (w_i k_j + w_jk_i) + \tau_{ij}, \nonumber
\eea
where $f_{\rho}$, $f_{v}$, $f_p$, and $f_{\pi}$ are random function 
of $\bk$; $\bw$ and $\bv$ are transverse vectors, 
$\bw \cdot \bk= \bv \cdot \bk= 0 $, and $\tau_{ij}$ is
a symmetric, traceless, transverse tensor, $\tau_i^i=\tau_{ij}k^j=0$.
The variables ($f_{\scriptsize \bullet}$), ($\bv$,$\bw$) and ($\tau_{ij}$) 
represent the scalar, vector, and tensor 
degrees of freedom of $T_{\mu\nu}^{(\sigma)}$ respectively. They are
the  source of the perturbation equations.

The goal  of the next three subsections is to express 
the correlators of the source
components $\SS^{(S)}$, $\SS^{(V)}$, and  $\SS^{(T)}$
in terms of these variables. These expressions, inserted in the
perturbation  equation~(\ref{diff}), then  allow us to  compute the CMB 
anisotropy and dark matter power spectra numerically.

\end{subsection}

%%%%%%%%%%%%%%%%%%%%%%%%%%%%%%%%%%%%%%%%%%%%%%%%%%%%%%%%%%%%%%%%%%%%%%%%%%%%%

\begin{subsection}{Axion seeds -- Scalar component}

We first consider the scalar contribution given
 by the four variables $f_{\scriptsize \bullet}$ of
Eq.\ \ref{Param}. Only two of these functions are independent, the other
two are related by energy and momentum conservation. We shall
use two linear combinations of  the three {\em scalar seed-functions}  
$f_\rho$, $f_v$, and $f_\pi$,
\bea
f_{\rho}({\bf k},\eta) &=& a^2\rho^{(\sigma)}
= T_{00}^{(\sigma)}({\bf k},\eta), \\ 
f_v({\bf k},\eta) &=& \frac{ik^jT^{(\sigma)}_{0j}({\bf k},\eta)}{k^2}, \\ 
f_{\pi}({\bf k},\eta)&=& \frac{3}{2k^4} 
\left[-T^{(\sigma)}_{ij}(\bk,\eta)k^ik^j 
+{1\over 3}k^2\delta^{kl}T^{(\sigma)}_{kl}(\bk,\eta)\right].
\eea

In the presence of seeds and in the linear perturbation approximation, 
the scalar component of the total
geometric perturbations determined by the Bardeen potentials 
$\Phi$ and $\Psi$ can be separated into a part
induced by the seeds, $\Psi_s$ and $\Phi_s$, given by
\be
k^2 \Phi_s = 4\pi G\left[f_{\rho} + 3 (\dot{a}/a)f_v)\right],
\,\,\,\,\,\,\,\,
\Phi_s + \Psi_s = - 8\pi G f_{\pi},
\label{Phiff}
\ee
and a part induced by the perturbations of the
cosmic fluid, $\Psi_m$ and $\Phi_m$.
The total geometric perturbations are given by the sums,
\be
\Psi = \Psi_s + \Psi_m, \,\,\,\,\,\,\,\,  \Phi = \Phi_s + \Phi_m.
\label{defsm}
\ee
The Bardeen potentials
are gauge invariant and fully describe scalar perturbations of the
Friedman  geometry (for details look in \cite{KS,DS}).

Scalar perturbations are seeded by $\Phi_s$ and $\Psi_s$. These are 
the standard
independent variables to  use  as scalar sources  in the perturbation
equations.  In order to simplify somewhat the computation, 
we  use $\Phi_s$ and $f_\pi$ as our scalar seed degrees of
freedom and the scalar source vector becomes
\be
\SS^{(S)}(\bk,\eta) = [\Phi_s(\bk,\eta) , 4\pi G f_{\pi}(\bk,\eta)].
\ee

The energy-momentum tensor of the axion is given by \Eq{Emt},
which leads to the following expressions for the seed-functions
in terms of the axion field $\si$:   
\bea
 f_{\rho}(\bk,\eta)&=&\frac{1}{2} 
\int \frac{d^3p}{(2 \pi)^3} 
\Big[ \dot{\sigma}(\bp,\eta) \dot{\sigma}(|\bk-\bp|,\eta)  
%\nonumber \\
-   \bp \cd
(\bk-\bp) \si(\bp,\eta) \si(|\bk-\bp|,\eta) \Big] , \\ 
 f_{v}(\bk,\eta)&=&-\frac{1}{k^2} 
\int \frac{d^3p}{(2 \pi)^3} \bk \cd
(\bk-\bp)
\dot{\sigma}(\bp,\eta) \sigma(|\bk-\bp|,\eta) , \\ 
 f_{\pi}(\bk,\eta)&=&-\frac{3}{2k^4}
\int \frac{d^3p}{(2 \pi)^3} 
\Big[ (\bk \cd \bp) [ \bk \cd (\bk-\bp)]  
%\nonumber \\ 
-  \frac{1}{3}k^2\bp \cd (\bk-\bp)
\Big] \sigma(\bp,\eta) \sigma(|\bk-\bp|,\eta).
\label{Fpi}
\eea
The first two seed-functions, $f_{\rho}$ and $f_v$,
together with \Eq{Phiff}, yield 
$\Phi_s$, 
\bea
\Phi_s(\bk,\eta)&=&\frac{4 \pi G}{k^2} 
\int \frac{d^3p}{(2 \pi)^3} 
\left[ \frac{1}{2}\dot{\sigma}(\bp,\eta) \dot{\sigma}(|\bk-\bp|,\eta)
\right. 
- \frac{1}{2}\bp \cd (\bk-\bp) \sigma(\bp,\eta)\sigma(|\bk-\bp|,\eta) 
\nonumber \\
&-& \left. 3 \frac{\dot{a}}{a} \frac{\bk \cd (\bk-\bp)}{k^2}
\dot{\sigma}(\bp,\eta) \sigma(|\bk-\bp|,\eta) \right].
\label{Phi}
\eea

The only information about the source random variables 
which we really need are the unequal time correlators 
between the Fourier components of the independent variables 
$\Phi_s$ and $f_{\pi}$. 
These correlators can be written in terms of four 
real (since the correlators
$\langle\si(\bk,\eta)\si^*(\bk',\eta')\rangle$ are real) 
{\em scalar source correlation functions}, $F_{11}$, $F_{22}$, 
$F_{12}$, and $F_{21}$, 
which completely characterize the scalar component of the source,
\bea
\langle \Phi_s(\bk,\eta) \Phi_s^*(\bkp,\eta') \rangle &=& 
\delta(\bk - \bkp) F_{11}(k,\eta,\eta') , \nonumber \\
4\pi G \langle \Phi_s(\bk,\eta) f_{\pi}^*(\bkp,\eta') \rangle &=& 
\delta(\bk - \bkp) F_{12}(k,\eta,\eta') ,\nonumber \\ 
4\pi G \langle f_{\pi}(\bk,\eta) \Phi_s^*(\bkp,\eta') \rangle &=& 
\delta(\bk - \bkp) F_{21}(k,\eta,\eta'),\nonumber \\
(4\pi G)^2 \langle f_{\pi}(\bk,\eta) f_{\pi}^*(\bkp,\eta') \rangle &=& 
\delta(\bk - \bkp) F_{22}(k,\eta,\eta'). \nonumber
\eea
Note that $F_{11}(k,\eta,\eta)$ and $F_{22}(k,\eta,\eta)$ are
positive by definition and, since the functions $F_{\bullet}$ are real,
$F_{ij}(k,\eta,\eta')=F_{ji}(k,\eta',\eta)$.
In order to compute these functions 
we make use of Eqs.\ (\ref{Fpi}) and (\ref{Phi}) and
we exploit the stochastic average conditions of the Gaussian variables
$\sigma$ and $\dot{\sigma}$ (Wick's theorem).
 We first introduce three real auxiliary variables
$\Sigma_1$, $\Sigma_2$, 
and $\Sigma_3$, which depend on the power spectrum of the axion field,
$\langle |c(\bk)|^2 \rangle$, and 
on the solution $\varphi$ of the evolution equation, \Eq{Evol2},  
\bea
\langle \sigma(\bk,\eta) \sigma(\bkp,\eta') \rangle &=&
(2\pi)^3 \delta(\bk - \bkp) \Sigma_1(k,\eta,\eta'), \nonumber \\
\langle \dot{\sigma}(\bk,\eta) \dot{\sigma}(\bkp,\eta') \rangle &=&
(2\pi)^3 \delta(\bk - \bkp) \Sigma_2(k,\eta,\eta'), \nonumber\\
\langle \sigma(\bk,\eta) \dot{\sigma}(\bkp,\eta') \rangle &=&
(2\pi)^3 \delta(\bk - \bkp) \Sigma_3(k,\eta,\eta'), \nonumber\\
\langle \dot{\sigma}(\bk,\eta) \sigma(\bkp,\eta') \rangle&=&
(2\pi)^3 \delta(\bk - \bkp) \Sigma_3(k,\eta',\eta).
\label{Sigma123}
\eea
The variables $\Si_i$ are given by
\bea
\Sigma_1(k,\eta,\eta')&=&\frac{\langle |c(\bk)|^2 \rangle}{ka(\eta) a(\eta')}
\varphi(k, \eta) \varphi(k, \eta'), \\
\Sigma_2(k,\eta,\eta')&=&\frac{\langle |c(\bk)|^2 \rangle}{ka(\eta) a(\eta')}
[ \dot{\varphi}(k, \eta)-{\cal H}(\eta) \varphi(k, \eta)]
[ \dot{\varphi}(k, \eta')-{\cal H}(\eta') \varphi(k, \eta') ],\\
\Sigma_3(k,\eta,\eta')&=&\frac{\langle |c(\bk)|^2 \rangle}{ka(\eta) a(\eta')}
[\dot{\varphi}(k, \eta)-{\cal H}(\eta)] \varphi(k, \eta'),
\eea
where ${\cal H}\equiv\dot{a}/a$. Notice that $\Sigma_1(\eta,\eta)$ and
$\Sigma_2(\eta,\eta)$ are positive by definition.

Inserting these results in Eqs.\ (\ref{Fpi}) and (\ref{Phi}),
and making use of Wick's theorem for the ``random variable'' $c(\bk)$,
we can  work out a somewhat lengthy but straight forward expression
for the scalar source functions, 
$F_{11}$, $F_{22}$, $F_{12}$, and $F_{21}$, in terms of the variables 
$\Sigma_1$, $\Sigma_2$, and $\Sigma_3$:
\bea
F_{11}&&(k,\eta,\eta')=\frac{(4\pi G)^2}{k^4} \int \frac{d^3p}{(2\pi)^3} 
\Big\{  
%\nonumber \\
\frac{1}{2} \Sigma_2(p,\eta,\eta')   \Sigma_2(|\bk-\bp|, \eta,\eta') 
\nonumber \\
&&-\frac{1}{2} \bp \cdot(\bk - \bp) 
\Big[ \Sigma_3(p,\eta,\eta') \Sigma_3(|\bk-\bp|, \eta,\eta') 
+ \Sigma_3(p,\eta',\eta) \Sigma_3(|\bk-\bp|, \eta',\eta) \Big] 
\nonumber \\
&&-3\frac{\bk \cdot(\bk - \bp)}{k^2} 
\Big[ {\cal H}(\eta) \Sigma_2(p,\eta,\eta') \Sigma_3(|\bk-\bp|, \eta,\eta')  
+ {\cal H}(\eta') \Sigma_2(p,\eta,\eta') \Sigma_3(|\bk-\bp|, \eta',\eta) \Big] 
\nonumber \\
&&+ \frac{1}{2} (\bp \cdot(\bk - \bp))^2 
\Sigma_1(p,\eta,\eta') \Sigma_1(|\bk-\bp|, \eta,\eta') 
+ 3 \frac{(\bp \cdot \bk - p^2)(k^2-\bp \cdot \bk)}{k^2} \times 
\nonumber \\
&& \ \ \ \Big[{\cal H}(\eta) \Sigma_3(p,\eta',\eta) \Sigma_1(|\bk-\bp|, \eta,\eta') 
+  {\cal H}(\eta') \Sigma_3(p,\eta,\eta') \Sigma_1(|\bk-\bp|, \eta',\eta) \Big] 
\nonumber \\
&&+9\frac{{\cal H}(\eta){\cal H}(\eta')}{k^4} \Big[ (\bk \cdot(\bk - \bp))^2 
\Sigma_2(p,\eta,\eta') \Sigma_1(|\bk-\bp|, \eta,\eta') 
\nonumber \\
&&+ (\bk \cdot(\bk - \bp))(\bk \cdot \bp) 
\Sigma_3(p,\eta',\eta) \Sigma_3(|\bk-\bp|, \eta,\eta') \Big] \Big\} , 
\nonumber \\
F_{22}&&(k,\eta,\eta') = \frac{9(4\pi G)^2}{2k^8 } \int \frac{d^3p}{(2\pi)^3}
\Big[ (\bk \cdot \bp)(\bk \cdot (\bk - \bp)) 
-\frac{1}{3} k^2 \bp \cdot (\bk -\bp) \Big]^2 
%\times 
%\nonumber \\
%&& \ \ \ 
\Sigma_1(p,\eta,\eta') \Sigma_1(|\bk-\bp|, \eta,\eta'), \nonumber \\
F_{12}&&(k,\eta,\eta')= -\frac{3(4\pi G)^2}{2 k^6}
\int \frac{d^3p}{(2\pi)^3}
\Big[ (\bk \cdot \bp)(\bk \cdot (\bk - \bp))
-\frac{1}{3} k^2 \bp \cdot (\bk -\bp) \Big] \times
\nonumber \\
&& \ \ \ \Big[ \Sigma_3(p,\eta',\eta) \Sigma_3(|\bk-\bp|, \eta',\eta) 
- \bp \cdot (\bk - \bp) \Sigma_1(p,\eta,\eta') \Sigma_1(|\bk-\bp|, \eta,\eta')
\nonumber \\
&&-6{\cal H}(\eta) \frac{\bk \cdot (\bk - \bp)}{k^2}
\Sigma_3(p,\eta',\eta) \Sigma_1(|\bk-\bp|, \eta,\eta') 
\Big],  \nonumber \\
F_{21}&&(k,\eta,\eta')=F_{12}(k,\eta',\eta) . \nonumber
\eea
The scalar source correlators of the perturbation
equation (\ref{diff}) can be written as a two by two positive 
and hermitian matrix, 
\be
\langle \SS_i^{(S)}(\bk,\eta) \SS_j^{(S)*}(\bk,\eta') \rangle 
= \left[
\begin{array}{cc}
F_{11}(k,\eta,\eta') & F_{12}(k,\eta,\eta') \\
F_{21}(k,\eta',\eta) & F_{22}(k,\eta,\eta')
\end{array}
\right].
\ee

\end{subsection}

%%%%%%%%%%%%%%%%%%%%%%%%%%%%%%%%%%%%%%%%%%%%%%%%%%%%%%%%%%%%%%%%%%%%%%%%%%%

\begin{subsection}{Axion seeds -- Vector component}

The vector contribution to the perturbation equations is seeded by 
the {\em vector seed-functions} $v_i$, \Eq{Param},
\be
v_i(\bk,\eta)= P_i^j T_{0j}^{(\si)}(\bk,\eta),
\ee
where $P_i^j$ is the projector operator onto the space 
orthogonal to $\bk$ defined by
\be
P_{ij}=\delta_{ij} - \hat{k}_i \hat{k}_j, \ \ \ \ \hat{k}_i=k_i/k. 
\ee
Again, the second vector seed function, $\bw$, is given by $\bv$ via
momentum conservations.
Defining the projection of the vector $\bp$ onto the space orthogonal
to $\bk$  by $\bp^{\bot}=P\bp$, we obtain an expression for 
the vector seed-functions in terms of the axion field, 
\be
v_j(\bk,\eta)=i\int\frac{d^3p}{(2\pi)^3}
p_{j}^{\bot}\dot{\si}(\bp,\eta)\si(\bk-\bp,\eta). 
\label{W}
\ee
We again need the unequal time correlators between the Fourier components
of the vector seed-functions $v_i$. These correlators 
can be written in terms of a {\em vector
source correlation function} $G$, which completely characterize the 
vector component of the source \cite{DKM},
\be 
(4\pi G)^2 \langle v_i(\bk, \eta) v_j(\bk, \eta') \rangle = (\delta_{ij} -
\hat{k}_i \hat{k}_j)G(k, \eta, \eta').
\ee
Using \Eq{W} and \Eq{Sigma123} this function takes the form
\be
G(k, \eta, \eta') = \frac{(4\pi G)^2}{2k^2} \int \frac{d^3p}{(2\pi)^3}
\left(k^2p^2-(\bk \cdot \bp)^2\right)  
\Big[\Sigma_2(p,\eta,\eta')  \Sigma_1(|\bk-\bp|,\eta,\eta')
+ \Sigma_3(p,\eta,\eta')  \Sigma_3(|\bk-\bp|,\eta',\eta)\Big].
\ee
The vector source correlators of the perturbation equation~(\ref{diff})
then are
\be
\langle \SS^{(V)}_i(\bk,\eta) \SS^{(V)}_j(\bk,\eta') \rangle=
 P_{ij} G(k, \eta, \eta').
\ee

\end{subsection}

%%%%%%%%%%%%%%%%%%%%%%%%%%%%%%%%%%%%%%%%%%%%%%%%%%%%%%%%%%%%%%%%%%%%%%%%%%%%%%

\begin{subsection}{Axion seeds -- Tensor component}

The tensor contribution to the perturbation equations is seeded by 
the {\em tensor seed-functions} $\tau_{ij}$, \Eq{Param},
\be
\tau_{ij}(\bk,\eta)=
\left(P_{i}^kP_{j}^n-\frac{1}{2}P_{ij}P^{kn}\right)T_{kn}^{(\si)}(\bk,\eta).
\ee
This leads to an expression for 
the tensor seed-function in terms of the axion field, 
\be
 \tau_{ij}(\bk,\eta)=-\int\frac{d^3p}{(2\pi)^3}
\left[p_{i}^{\bot}p_{j}^{\bot}-(1/2)(\delta_{ij}-\hat{k}_i \hat{k}_j)
(p^{\bot})^2\right]    
\si(\bp,\eta)\si(\bk-\bp,\eta), 
\label{Tau}
\ee
which can be used to 
compute the unequal time correlators. 
These correlators 
can be written in terms of a {\em tensor
source correlation function}, $H$, which completely characterizes the 
tensor component of the source \cite{DKM}, 
\bea
(4\pi G)^2 \langle \tau_{ij}(\bk, \eta) 
\tau_{lm}(\bk, \eta') \rangle &=&[\delta_{il}\delta_{jm}+
\delta_{im}\delta_{jl}-\delta_{ij}\delta_{lm}+
k^{-2}(\delta_{ij}k_lk_m + \delta_{lm}k_ik_j - 
\delta_{il}k_jk_m  \nonumber \\
&-& \delta_{im}k_lk_j - \delta_{jl}k_ik_m  - 
\delta_{jm}k_lk_i )+
k^{-4}k_ik_jk_lk_m] H(k,\eta,\eta')  \nonumber \\
&=&  \left(P_{il}P_{jm}+ P_{jl}P_{im} -P_{ij}P_{lm}\right)  
     H(k, \eta, \eta'). 
\label{Tens}
\eea
Using \Eq{Tau} and \Eq{Sigma123} this function takes the form
\bea
H(k, \eta, \eta') &=& \frac{(4\pi G)^2}{4k^4} \int \frac{d^3p}{(2\pi)^3}
\left(k^2p^2-(\bk \cdot \bp)^2 \right)^2  
\Sigma_1(p,\eta,\eta')  \Sigma_1(|\bk-\bp|,\eta,\eta').
\label{Tens12}
\eea
The tensor source correlators of the perturbation equation, \Eq{diff}, 
hence are
\be
\langle \SS^{(T)}_{ij}(\bk,\eta) \SS^{(T)}_{lm}(\bk,\eta') \rangle=
  \left(P_{il}P_{jm} + P_{jl}P_{im}- P_{ij}P_{lm}\right)  
     H(k, \eta, \eta').
\ee
\end{subsection}

\end{section}
\vspace{1cm}

%%%%%%%%%%%%%%%%%%%%%%%%%%%%%%%%%%%%%%%%%%%%%%%%%%%%%%%%%%%%%%%%%%%%%
%%%%%%%%%%%%%%%%%%%%%%%%%%%%%%%%%%%%%%%%%%%%%%%%%%%%%%%%%%%%%%%
%%%%%%%%%%%%%%%%%%%%%%%%%%%%%%%%%%%%%%%%%%%%%%%%%%%%%%%%%%%%%%%%%%%%%%%

\begin{section}{CMB anisotropies induced by
axion seeds}

In this section we present the CMB power spectrum
obtained in our scenario. We first describe the CMB angular power
spectrum obtained in
the coherent approximation and in Subsection~B we then show in
detail that the coherent approximation is very good for axionic seeds,
leading to errors of 5\% or less.

\begin{subsection}{CMB power spectrum -- Coherent approximation}
\label{CMBcoe}

A source is called coherent \cite{ds,pedro} if the 
unequal time correlation functions can be factorized or
replaced by the product of deterministic sources, as in \Eq{Coerente},
\be
\langle \SS_j(\eta) \SS_i(\eta') \rangle \simeq
\pm \sqrt{\langle |\SS_j(\eta)|^2\rangle\langle |\SS_i(\eta')|^2\rangle}.
\label{Ccoe}
\ee
As pointed out in 
Subsection~\ref{source}, 
this approximation is exact only if the source evolution is linear. 
Then the different $\bk$ modes do not mix and the value of the 
source term at a fixed $\bk$ at a later time is given by 
its value at initial time multiplied
by some transfer function, as in \Eq{Transfert}.
In this situation \Eq{Ccoe} becomes an equality and the model
is perfectly coherent.
This is not the case for our model since we know that, although the 
axion field evolves according to a linear equation, 
its energy-momentum tensor, 
which  enters into the perturbation equations as source, does not; it
is quadratic in the field $\si$.
Thus, nonlinearity leads to mixing of scales and 
to deviation from a Gaussian distribution.

Nevertheless our situation is very similar to the large $N$
limit of global $O(N)$ models in which the only nonlinearities also
are the quadratic expressions of the energy-momentum tensor. In this
case the effects of decoherence are very small
and one finds that the full incoherent result is not very 
different from the perfectly coherent approximation \cite{DKM}.

This result motivated us to compute the CMB anisotropy
in the perfectly coherent approximation. Here we repeat and expand on
results already presented in \cite{afrg} while in the next subsection 
we justify them by discussing the full incoherent case.

\begin{figure}[ht]
\centerline{\epsfig{file=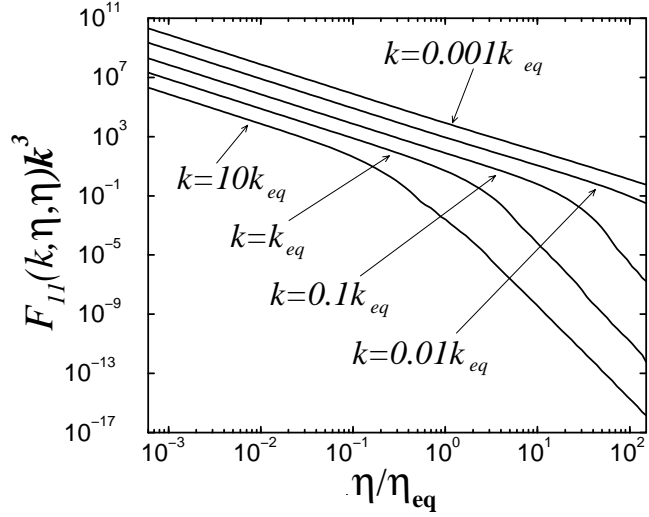, width=3.5in}}
\caption{Time evolution of the source function $F_{11}(k,\eta,\eta)k^3$, 
with  tilt $n_{\si}=1.1$, for different modes, $k=0.001k_{eq}$, 
$k=0.01k_{eq}$, $k=0.1k_{eq}$, $k=k_{eq}$, and $k=10k_{eq}$.
For super-horizon modes, the correlator $F_{11}$
decays like $\eta^{1-2n_{\si}}/k^4$.
As soon as a mode enters the horizon the corresponding correlator 
decays faster due to the oscillating behavior of the axion field.
Before crossing the horizon, the other scalar equal time 
correlators show the same power law behavior while the 
vector correlator
$G(k,\eta,\eta) \propto \eta^{1-2n_{\si}}/ k^2$ and
the tensor correlator
$H(k,\eta,\eta) \propto \eta^{1-2n_{\si}}$ (independent of $k$).}
\label{modesint}
\end{figure}

In order to compute the CMB anisotropy power spectrum in the coherent
approximation, we replace the unequal time correlation functions  
in \Eq{pow} by the products
\bea
\langle \SS_i^{(S)}(\bk,\eta) \SS_j^{(S)}(\bk,\eta') \rangle&=&
F^{(n_{\si})}_{ij}(k,\eta,\eta') \simeq
\pm[F^{(n_{\si})}_{ij}(k,\eta,\eta)F^{(n_{\si})}_{ij}(k,\eta',\eta')]^{1/2},
\nonumber \\
\langle \SS^{(V)}(\bk,\eta) \SS^{(V)}(\bk,\eta') 
\rangle &=& G^{(n_{\si})}(k,\eta,\eta') \simeq
[G^{(n_{\si})}(k,\eta,\eta)G^{(n_{\si})}(k,\eta',\eta')]^{1/2},\\
\langle \SS^{(T)}(\bk,\eta) \SS^{(T)}(\bk,\eta') 
\rangle &=& H^{(n_{\si})}(k,\eta,\eta')\simeq
[H^{(n_{\si})}(k,\eta,\eta)H^{(n_{\si})}(k,\eta',\eta')]^{1/2}, \nonumber
\eea
where we have indicated the dependence of the correlators
on the spectral index $n_{\si}$ by a super-script.
In Fig.~\ref{modesint} we show the time behavior of one of the 
equal time correlators. On super-horizon scales, $k\eta\ll 1$, they
all  display the same typical behavior, 
$\propto k^{-\kappa} \eta^{1-2n_{\si}}$, which
depends on the spectral index $n_{\si}$ and on $\kappa$, a positive 
power determined by dimensional arguments.
On sub-horizon scales the correlators  decay fast due to incoherent
oscillations of the convolved axion field.

We have solved \Eq{diff} for the scalar, vector, and tensor components.
The CMB anisotropy power spectrum is given by the sum of the three 
contributions and depends on the spectral index $n_{\si}$,
\be
C^{(n_{\si})}_{\ell}=C^{(Sn_{\si})}_{\ell}+C^{(Vn_{\si})}_{\ell}
+C^{(Tn_{\si})}_{\ell}.
\ee

\begin{figure}[ht]
\centerline{\epsfig{file=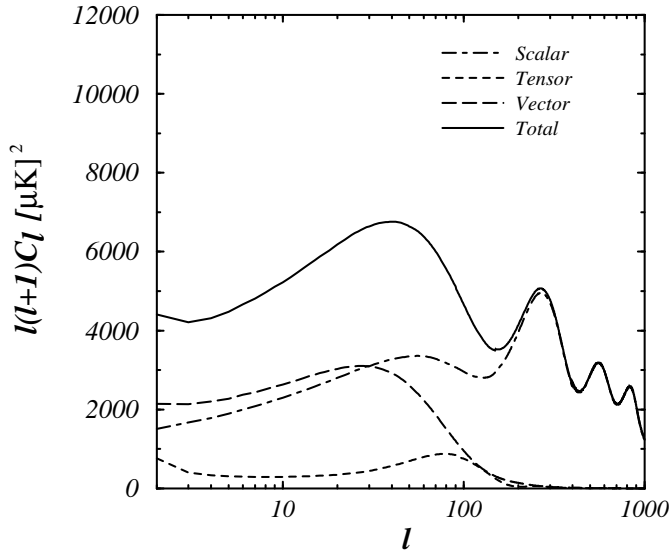, width=4in}}
\caption{The CMB anisotropy power spectrum for fluctuations induced 
by axion seeds with a tilt $n_{\si}=1.1$ and $\Lambda=0$. This
result is computed within the coherent approximation. We show the 
scalar (dot-dashed), vector (dashed) and 
tensor (dotted) contributions separately as well as their sum (solid).}
\label{svtcl}
\end{figure}

In Fig.~\ref{svtcl} we show the scalar, vector, and
tensor contributions to the resulting CMB anisotropies for an axion
spectrum with tilt $n_{\si}=1.1$. The ``hump'' at $\ell\sim 60$ 
in the scalar component is due to
the isocurvature nature of the perturbations. This is also one of the 
reasons why the acoustic peaks are very low, 
the other being that the vector (and tensor) component is of the same
order of magnitude as the scalar one. This
enhances, in seed models, the CMB spectrum at large scales thereby lowering 
the acoustic peaks at small scales. The result obtained is remarkably
similar to the large $N$ case studied in \cite{DKM}. The main
difference here is that, like for usual inflationary models,  we 
dispose of a spectral index which is basically free. By
choosing slightly bluer spectra, we can enhance the power on
smaller scales. 

\begin{figure}[ht]
\centerline{\epsfxsize=4in  \epsfbox{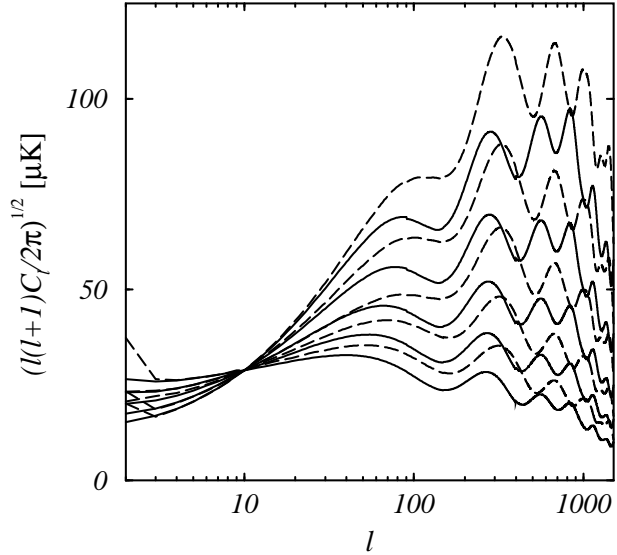}}
\caption{The CMB anisotropy power spectrum for fluctuations induced 
by axion seeds. We show the sum of the scalar, vector,  and 
tensor contributions for 5 different tilts, with 
$\Omega_{\Lambda}=0$ (solid) and $\Omega_{\Lambda}=0.7$
(long dashed). The tilt is raising from bottom to top,
 $n_{\si}=1.1,~1.2,~1.3,~1.4,~1.5$.}
\label{cl}
\end{figure}

In Fig.~\ref{cl} we show the sum of the scalar, vector, and tensor
contributions comparing the results from different tilts
with and without a
cosmological constant. The CMB power spectra obtained 
can have considerable acoustic peaks at $\ell \sim 250$ to $300$,
which can be raised further by adding a 
non-vanishing cosmological constant.
Increasing the tilt $n_{\si}$ raises the acoustic peaks and moves them
to slightly smaller scales. 
As found in \cite{1}, the power spectrum of the scalar component 
 is always blue.  The tensor and
 vector component counterbalance the increase of the tilt,
 maintaining a nearly scale invariant spectrum on large scales.
The models can be clearly
discriminated from the common inflationary spectra by their 
isocurvature hump and by the position of the first peak.
A discussion on the comparison of these results with recent CMB data
will be given in Section~\ref{cmbdata}.

We have also computed the CMB polarization for our model. The result
for two different spectral indices is shown in Fig.~\ref{polar} where
we compare it with the polarization from usual inflationary models.
It is interesting to note that our models show a characteristic
``polarization hump''  which is  significantly smaller in inflationary
models.  The polarization ``hump'' is completely  suppressed for
topological defects due to causality~\cite{HSW} and
represents  a very  characteristic signature of ``acausal seed models'' 
like the one under  consideration.

\begin{figure}[ht]
\centerline{\epsfig{file=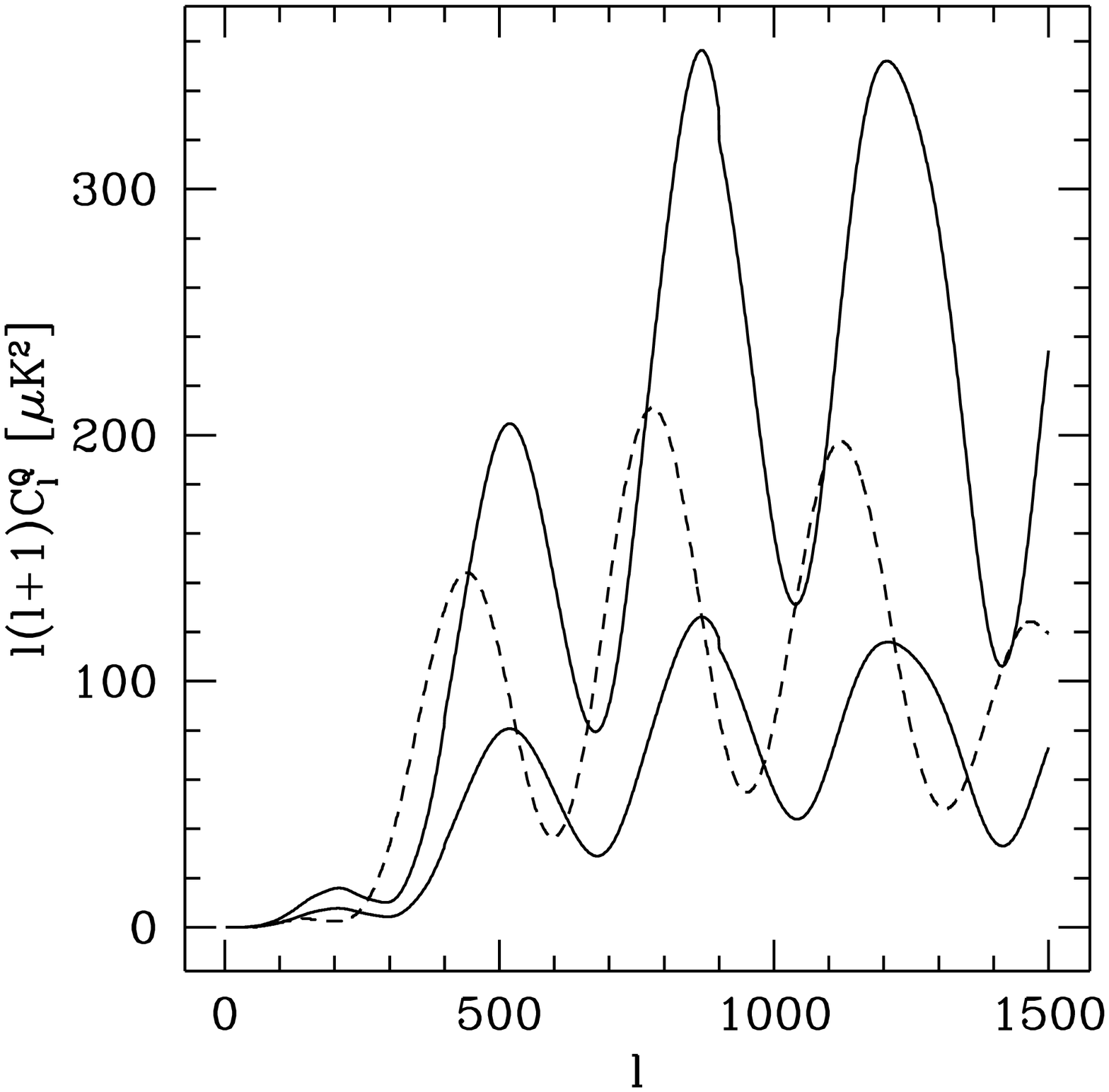, width=3in}
                    \epsfig{file=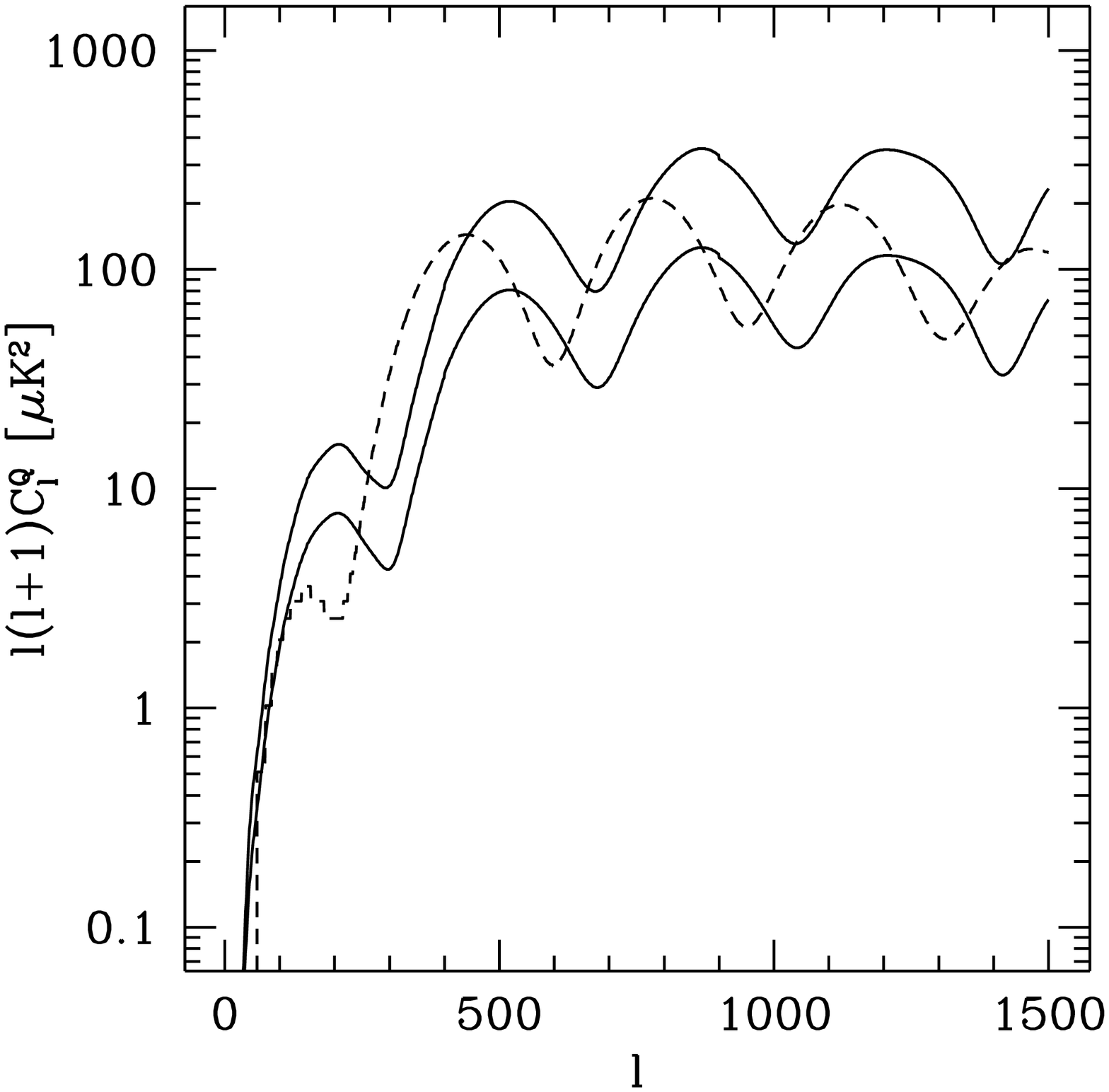, width=3in} }
\caption{The CMB polarization power spectrum in linear scale (left)
and log scale (right) for fluctuations induced 
by axion seeds and shown for 2 different tilts, with 
 $\Omega_{\Lambda}=0.7$, $n_\si=1.3$ (lower solid line) and $n_\si=1.5$
(upper solid line), are compared with the standard inflation 
result (dashed line) for the same cosmological
parameters. Polarization clearly distinguishes between
inflation and axion seeds, especially via the isocurvature hump.}
\label{polar}
\end{figure}

\end{subsection}

%%%%%%%%%%%%%%%%%%%%%%%%%%%%%%%%%%%%%%%%%%%%%%%%%%%%%%%%%%%%%%%%%%%%%%%%%%%%%%

\begin{subsection}{Decoherence}
\label{deco}

In order to estimate the accuracy of the results found in the 
previous subsection, we discuss here the decoherence of the
axion seeds showing that the difference between the coherent 
approximation and the full incoherent calculation is very small.
The decoherence is tested  only for the scalar
component of the spectrum, where it may lead to ``smearing out'' of the
acoustic oscillations. Its effects on vector and tensor perturbations
are expected to be small.

We first introduce the property of ``scaling'' for the
axion seeds. When working with seeds,
to solve the  problem of the enormous dynamical range\footnote{To compute
the CMB and dark matter power spectra, we need to know the seed
functions over a dynamical range of $k_{\max}/k_{\min}\sim 30'000$ and
this for all times $\eta_{in} \le \eta,\eta' \le \eta_0$ with
$k\eta_{in} \ll 1$. This gives finally more than 1000 functions of two
variables which have to be known accurately over a long time
interval.} needed to compute the $C_\ell$'s from $\ell=2$ to
$\ell=1500$,  one  often makes use of scaling properties.
We call seeds scaling if their correlation function,
$\langle \SS(\bk,\eta) \SS(\bk,\eta')    \rangle$,
is scale free, {\it i.e.}, the only dimensional parameters in $F_{ij}$, 
$G$, and $H$ are the
variables $\eta$, $\eta'$, and $\bk$ themselves. As we have already
mentioned, axion seeds are not scaling since the correlation function
contains factors of the form $(k/k_1)^{\alpha}$. But such a simple
pre-factor can we written as 
\[ (k/k_1)^{\alpha} = (k\eta)^{\alpha}/(k_1\eta)^{\alpha} \]
and does not enter the costly numerical integration.
Numerical calculations are reduced greatly if one can write the
correlation function in the form 
\bea
F_{ij}(\bk,\eta,\eta') &=& f(\sqrt{\eta\eta'},k_1) 
C_{ij}(y,r), \nonumber \\
G(\bk,\eta,\eta') &=& g(\sqrt{\eta\eta'},k_1) W(y,r), 
\label{Scaling}  \\
H(\bk,\eta,\eta') &=& h(\sqrt{\eta\eta'},k_1) T(y,r),  \nonumber
\eea
where $y \equiv k\sqrt{\eta \eta'}$ and $r \equiv \sqrt{\eta'/ \eta}$, and 
$f$, $g$, and $h$ are given explicitly. The matrix $C_{ij}$ and the 
functions $W$ and $T$ are dimensionless by construction.
In the following we shall call this behavior ``modified scaling''.

But even after this extraction of the explicit breaking of scaling,
our source does not exactly obey ``modified scaling'' due to the
radiation-matter transition. As one can see
immediately from the evolution 
equation of the axions in the post-big bang phase, \Eq{Evol2}, the extra 
dimensional parameter implicitly contained in the unequal time correlators
is $\eta_{*}$ which comes from the expression for
the scale factor $a$, \Eq{Effpot}. The radiation-matter transition
introduces the new scale $\eta_{*}$ and thereby spoils the modified scaling 
behavior of the axion seeds\footnote{This breaking of scale invariance
is also found in models with topological defects.}. However, deep in
the radiation or matter era, $\eta\ll\eta_*$ or $\eta\gg \eta_*$
respectively, the reduced correlation functions do obey scaling. In
order to avoid this  problem and to simplify the
numerical calculations, we therefore compute the axion field according to 
the equation for the pure radiation era, {\it i.e.}, 
setting $a(\eta)=\eta$. We call this the
{\em  radiation approximation}. This approximation affects   
the correlators and the CMB anisotropy power spectrum, especially at large
angular scales, but is expected not to differ significantly from the
correct results on the scales of the acoustic peaks, and it allows us
to  obtain sources which obey modified scaling.

In the coherent case, where we just need the equal time correlators,
the numerical requirements have not been very involved and we have
not been pushed to the radiation approximation. But, as we
shall see, the fully decoherent calculation will not change the
results considerably and therefore an enormous numerical effort, which
would be needed to compute the unequal time correlators without any
use of scaling behavior, is not justified for this simple test.

In the matter dominated era, axion seeds are amplified 
by quantum particle creation while in the radiation 
approximation they do not experience this amplification.
Nevertheless, axions are massless particles and they behave like a perfect 
radiation fluid. 
Thus, their energy density decreases as $1/a^4$, 
faster than the cosmic fluid in a matter dominate universe, 
where $a\propto \eta^2$ and $\rho\propto a^{-3}$, than in a radiation
dominated universe, where $a \propto \eta $ and $\rho\propto a^{-4}$.     
This leads one to some overestimation of   
the sources at $\eta > \eta_*$ in the radiation approximation.

\begin{figure}[ht]
\centerline{\epsfig{file=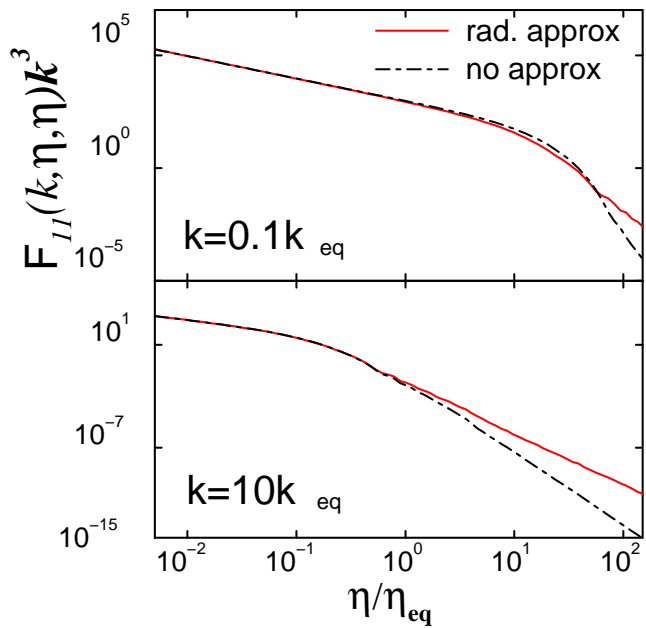, width=4in}}
\caption{Time behavior of $F_{11}(k,\eta,\eta)k^3$,
with spectral index $n_{\si}=1.1$, 
for a mode which enters the horizon before 
matter-radiation equality, $\varphi(k=10k_{eq},\eta)$,
and after, $\varphi(k=0.1k_{eq},\eta)$. Solid lines show the modes in
the radiation approximation, dashed lines 
without approximation. For $k>k_{eq}$ there is no difference on
super-horizon scales, while for $k<k_{eq}$ the additional
amplification experienced in the matter dominated phase is lost in the
radiation approximation. On sub-horizon scales, the radiation
approximation decays slower than the correct result.
A similar behavior is found for the other correlators.}
\label{radsource}
\end{figure}

In Fig.~\ref{radsource} we compare the time behavior
of one of the equal time correlators taking into account the
radiation matter transition (dashed) with those obtained in the
radiation approximation (solid line)
for two different values of $k$.
Modes that enter the horizon before matter-radiation equality,
$k > k_{eq}$, do not feel quantum particle creation; therefore, there is
no difference between the full result and the radiation approximation
on super-horizon scales. Inside the horizon, in the matter era the
mode decays faster than in the radiation approximation.
Modes which enter the horizon after equality,
$k < k_{eq}$, get first amplified by particle creation, an effect
which is missed in the radiation approximation, but
then decay faster than in the radiation approximation. 
As can be seen in Fig.~\ref{clradmat}, the slower decay has 
consequences on the CMB anisotropy power spectrum: using the radiation 
approximation somewhat enhances  the Sachs-Wolfe plateau and the first peak.
 
\begin{figure}[ht]
\centerline{\epsfig{file=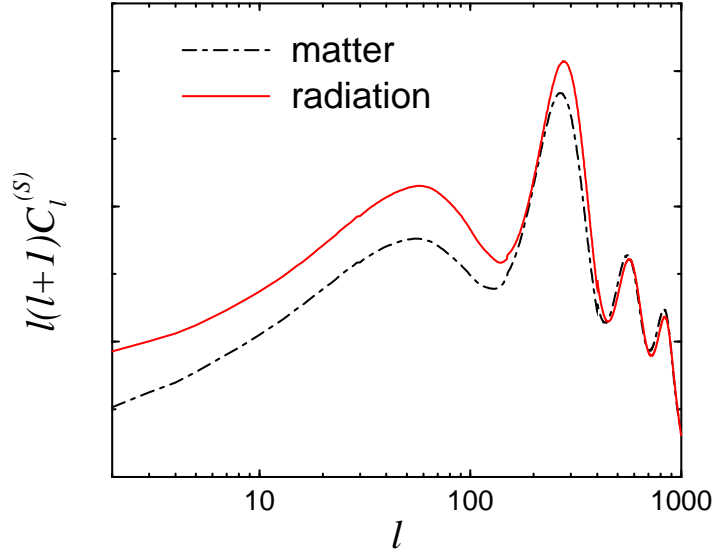, width=4in}}
\caption{Scalar contribution to the CMB angular power spectrum, computed 
in the pure radiation approximation (solid line) and without 
approximation (dashed line),
with an axion spectral index $n_{\si}=1.1$ and $\Lambda=0$.}
\label{clradmat}
\end{figure}

We now compute the  CMB anisotropies in the full decoherent case
for  the radiation approximation, making use of modified
scaling. We restrict our attention to the scalar 
component, where decoherence can be important.
 
As explained in \Eq{Scaling} we write the scalar correlation matrix
$F_{ij}$ (for $n_\si=1$) as
\be
F_{ij}(\bk,\eta,\eta')=(\eta \eta')^{3/2}C_{ij}(y,r),
\ee
where $C_{ij}$ is only function of $y$ and $r$ and hence dimensionless.
The matrix $C_{ij}$ is clearly symmetric under $r \rightarrow 1/r$ as 
can be seen in Fig.~\ref{sr}.
For $y <1$ the sources decay like $1/y$ and after horizon crossing they
begin to decay faster due to oscillations.

\begin{figure}[ht]
\centerline{\epsfig{file=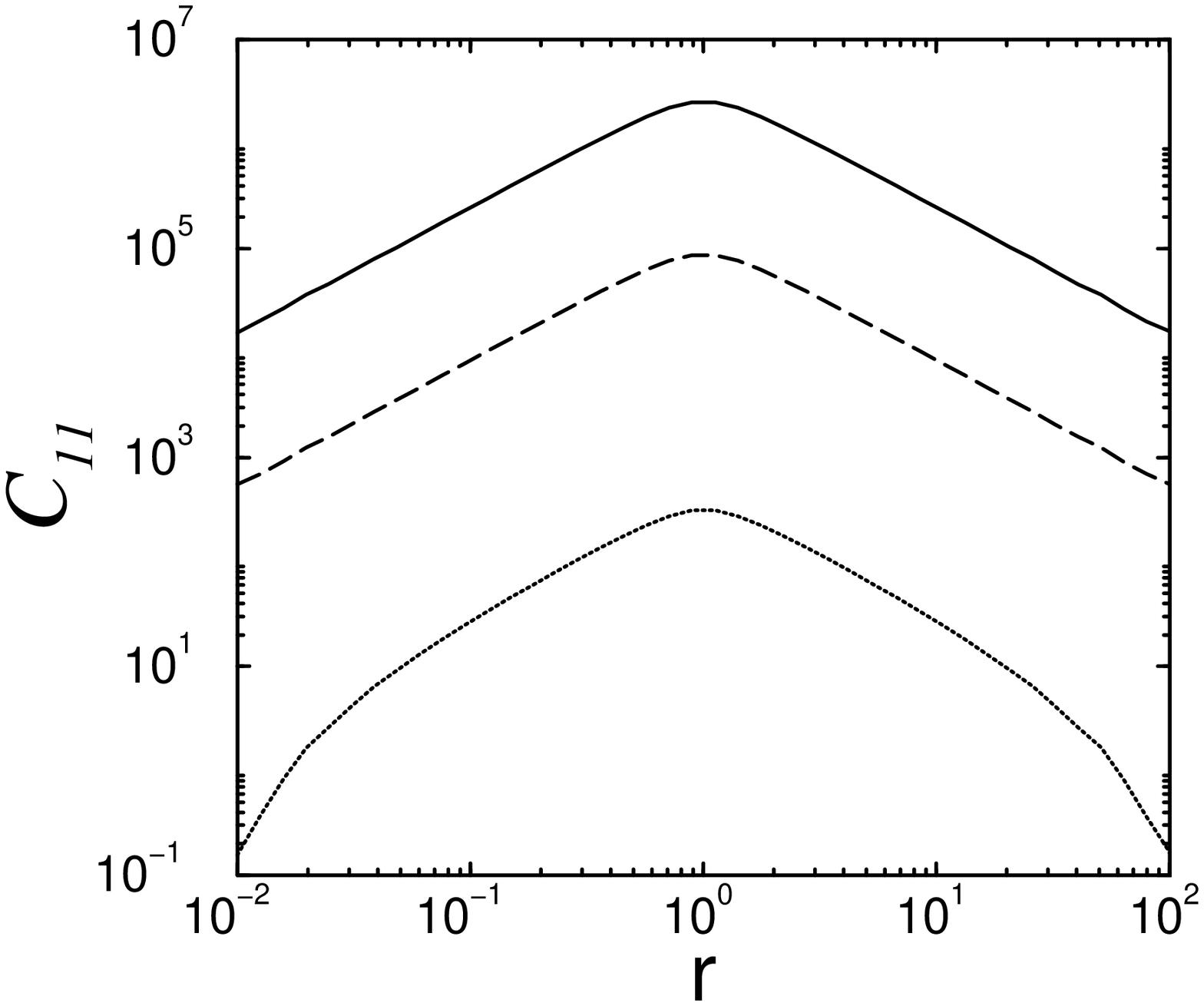, width=3in} 
\epsfig{file=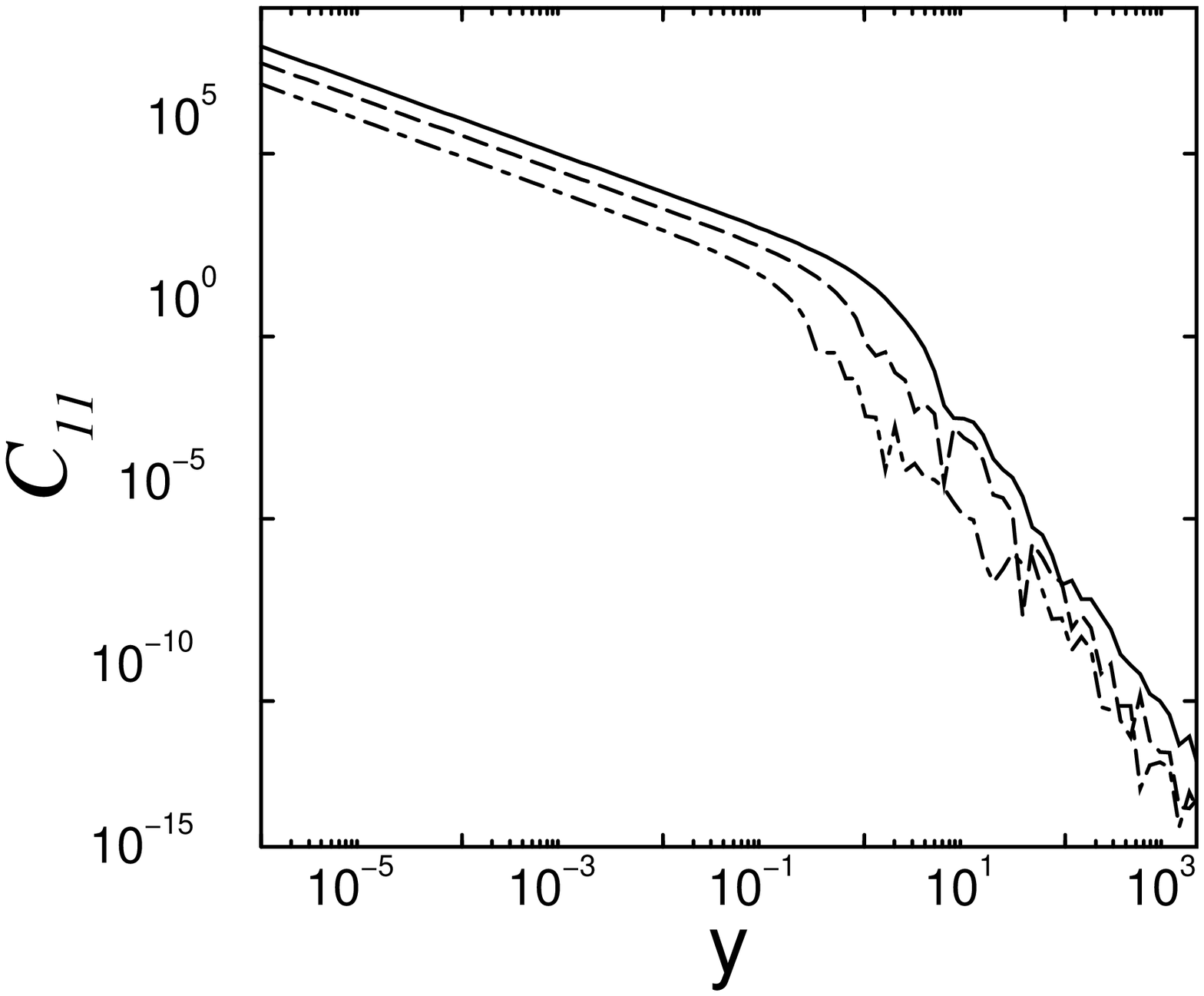, width=3in}}
\caption{The correlator $C_{11}(y,r)$ is shown. In the 
left panel the solid, dashed, and dotted lines respectively represent 
$C_{11}(1 \times 10^{-7},r)$, $C_{11}(1 \times 10^{-4},r)$, 
and $C_{11}(0.03,r)$. 
In the right panel the solid, dashed, and dotted lines 
respectively represent 
$C_{11}(y,1)$, $C_{11}(y,0.3)$, and $C_{11}(y,0.1)$. 
The other scalar correlators $C_{22}$ and $C_{12}$ behave similarly.}
\label{sr}
\end{figure}

The source correlation matrix $C_{ij}$ 
can now be considered as kernel of a positive hermitian 
operator in the variables $x=k\eta=y/r$ and $x'=k\eta'=yr$, which can be 
diagonalized \cite{DKM},
\be
C_{ij}(x,x')=\sum_{n} \lambda_n v_{in}(x) v_{jn}(x'),
\label{Sumeigs}
\ee
where $\{v_{in}\}$ is an orthonormal series of eigenvectors 
(ordered according to the amplitude of the corresponding eigenvalues) of 
the operator $C_{ij}$ for a given weight function $w$.
The eigenvectors and the eigenvalues depend on the weight
function $w$ which can be chosen to optimize the speed 
of convergence of the sums (\ref{Sumeigs}). 

Inserting \Eq{Sumeigs} in \Eq{pow} leads to 
\be
\langle X_i(\bk,\eta_0)X_j(\bk,\eta_0)\rangle=
\sum_n \la^{(n)}X_i^{(n)}(\bk, \eta_0) X_j^{(n)}(\bk, \eta_0),
\ee
where $X_i^{(n)}(\eta_0)$ is the solution of \Eq{diff} 
with deterministic source term $v_i^{(n)}$,
\be
X_j^{(n)}(\bk,\eta_0)=
\int_{\eta_{in}}^{\eta_{0}} d \eta \GG(\bk,\eta_0,\eta)_{jl}v_l^{(n)}(\bk,x).
\ee
For the scalar CMB anisotropy spectrum this gives
\be
C_{\ell}^{(S)}=\sum_{n=1}^{N} \lambda_n^{(S)} C_{\ell}^{(Sn)};
\label{Clsdeco}
\ee
$C_{\ell}^{(S)}$ is the scalar component of the CMB anisotropy induced by the 
deterministic source $v_n$ and $N$ is the number of eigenvalues which
have to be considered to achieve good accuracy.

In our model we actually find it easier to diagonalize the matrix
\[
\tilde{C}_{ij}(x,x')=C_{ij}(x,x')\sqrt{x x'},
\]
whose diagonal is flat for $x < 0.01$, exactly as
in the large-$N$ and texture models studied
in \cite{DKM}. In this case we have
\be
C_{ij}(x,x')=\sum_{n}^N \tilde{\lambda}_n 
\frac{\tilde{v}_{in}(x)}{\sqrt{x}}
\frac{\tilde{v}_{jn}(x')}{\sqrt{x'}},
\ee  
where ${\tilde{v}_{jn}}$ and ${\tilde{\lambda}_n}$ 
are the eigenvectors and the eigenvalues of 
the matrix $\tilde{C}_{ij}$.

\begin{figure}[ht]
\centerline{\epsfig{file=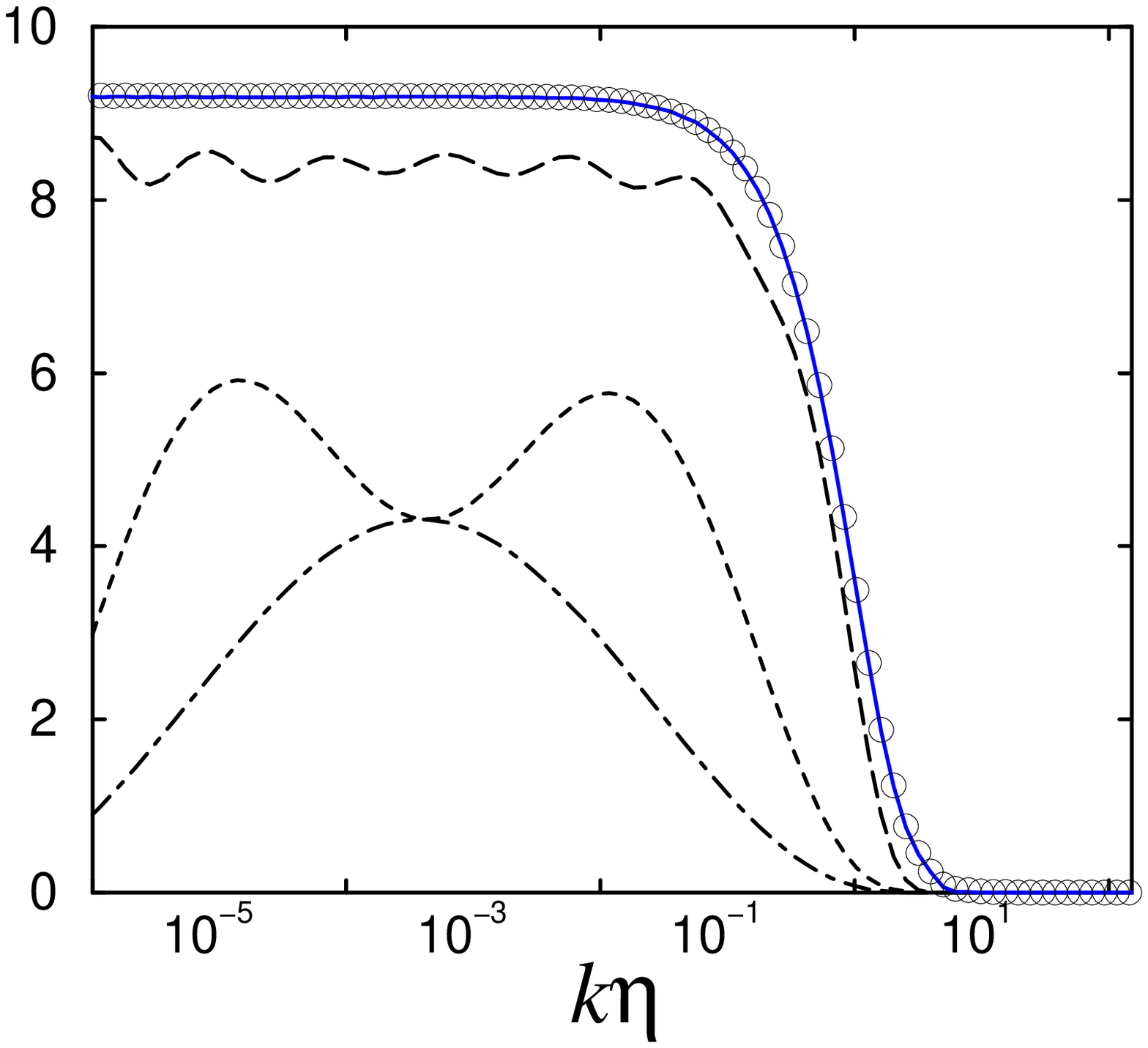, width=4in}}
\caption{The sum of the first few eigenfunctions of
$\tilde{C}_{11}(x,x')$ is shown for a weight function 
$w=1/x$. The first (dot dashed), first and second (short dashed),
first ten (long dashed), and fifty (solid) eigenfunctions
are summed up. The open circles represent the full correlation function.  
Here we only show the equal time diagonal of the correlation 
matrix but the same convergence behavior is found 
in the $C_{\ell}$ power spectrum which is sensitive to the full
correlation matrix.}
\label{diago}
\end{figure}

We diagonalize the matrix $\tilde{C}_{ij}$ using 
the logarithmic weight function $w=1/x$ which 
allows us to sample  the range of scales of interest more evenly. 
In Fig.~\ref{diago} we show the eigenvectors 
decomposition of one of the scalar correlators. 
Note that a rather high number of eigenvectors and eigenvalues 
is required to reach 
a good accuracy in the approximation of the diagonal of the
correlation function.
Summing up $N=50$ eigenvectors the convergence is guaranteed; the
summed up correlation function reproduces the original to better than $1\%$.

This is different from the large-$N$ model, where about 20 eigenvectors
suffice for the same accuracy. We assume that this difference is due
to the slower decay of the source functions. As can be seen from
Fig.~\ref{diago}, the source
function is decaying from its original value to about 1\% over the
interval  $0.1<k\eta <10$, while in the large-$N$ model this decay is
achieved in the interval   $0.5 <k\eta <4 $.

\begin{figure}[ht]
\centerline{\epsfig{file=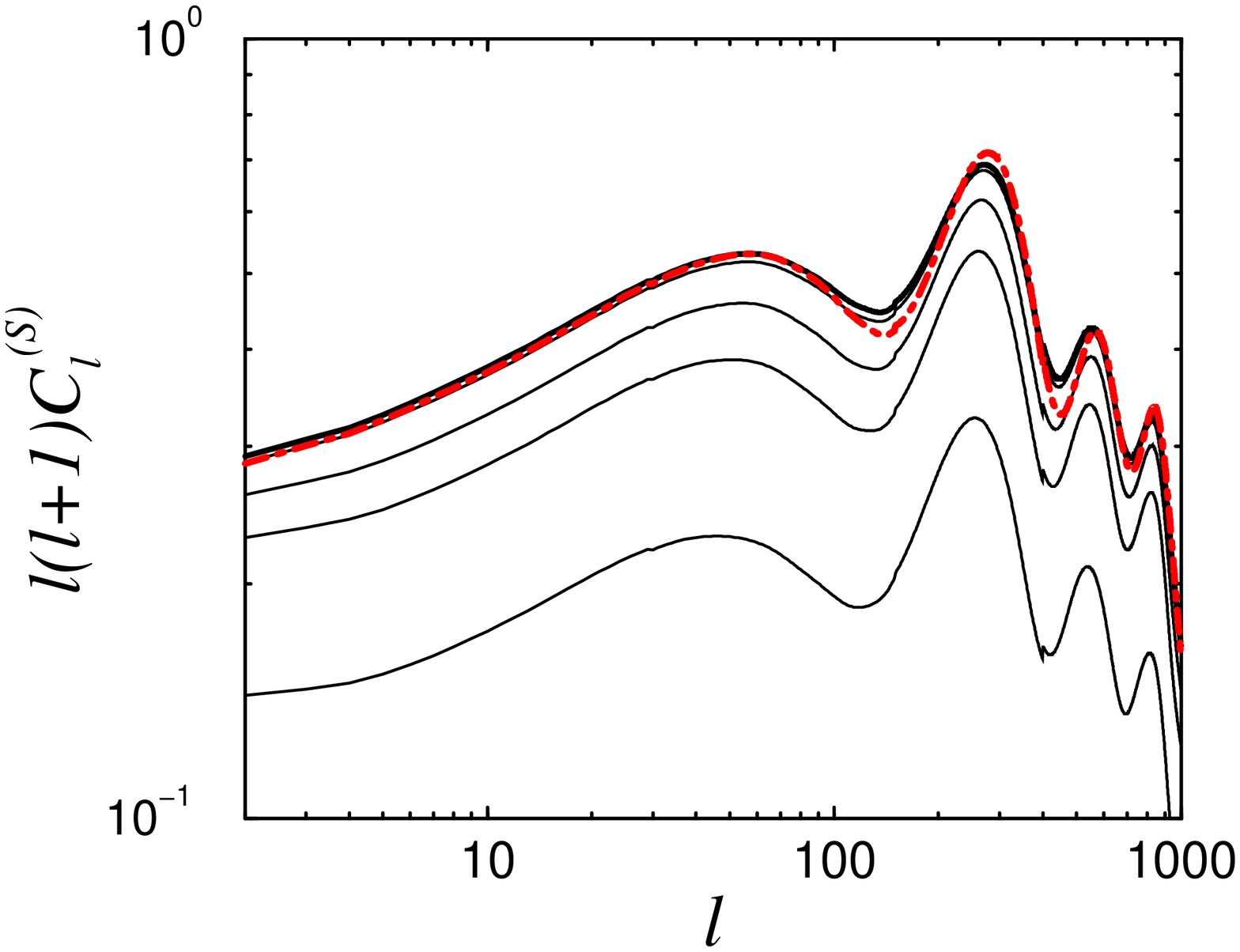, width=4in}}
\caption{The scalar contribution to the $C_{\ell}^{(S)}$ power spectrum
is shown for a primordial spectral index $n_{\si}=1.1$. From bottom to top, the
solid lines show the contributions of the sum of the first ten, 
first twenty, first thirty and first forty $C_{\ell}^{(Sn)}$'s.
The thick solid line represents the full eigenvectors summation (up to $N=50$)
to be compared to the perfect coherent approximation, 
shown by the dashed line. The decoherence does not significantly
wash out the acoustic peak and the oscillations.}
\label{cldeco}
\end{figure}
\end{subsection}

We now compute the scalar contribution to the CMB anisotropies  
using \Eq{Clsdeco}. The result is shown in Fig.~\ref{cldeco}.  
We note that decoherence slightly reduces the amplitude
of the oscillations around the first peak 
leaving however the secondary peaks and their 
positions almost unaffected. 
Although axion perturbations are in principle incoherent,
it is difficult to observe this from the CMB power spectrum.
The effects of decoherence are indeed very weak
and the spectrum obtained in the perfect coherent
approximation reproduces the decoherent result within less than 
$5\%$.
We hence are confident to obtain a sufficient accuracy in the 
perfectly coherent approximation which we shall apply 
for the rest of  this paper.

\end{section}
\vspace{1cm}

\begin{section}{Comparison with CMB anisotropy data and matter perturbations}
\label{cmbdata}
In this section we compare the results found in the previous section 
with data discussing in particular
the consequences of the normalization of CMB anisotropies to COBE scale
and presenting the cosmological parameters favored by our model.  
In Subsection~D we finally compute the dark matter power spectrum and
we compare it with data. 

\begin{subsection}{Normalization and the kink}
Comparing our numerical result with the CMB data we normalize our
curve to the fluctuation amplitude observed by COBE. This provides
a relation between the string and the scale of the break
$k_b$. Since we  ignore constant factors of order unity in the overall
amplitude in our calculation, the result for the amplitude is
not very precise, but certainly correct within a factor of about $2$.
For the best fit value of the tilt, $n_{\sigma} -1=\varepsilon \sim 0.33$, 
our numerical
result on the COBE scale (at $\ell \sim 10$) is 
 $\ell(\ell+1)C_\ell\simeq 0.3g_1^4(\eta_*k_b)^{-2\varepsilon}$. Here $g_1$ is
the dimensionless string coupling constant given by $\om_1/m_{\rm
Planck}$ where $\om_1=k_1/a_1=H(\eta_1)$ is the inverse string scale. 
Comparing this with the COBE normalization,
$\ell(\ell+1)C_\ell T_0^2 \simeq 5225\mu$K$^2$, yields
\be
\eta_*k_b =  (2.1\times 10^3g_1^2)^{1/\varepsilon}. \label{norm}
\ee
For example, if the string scale is $10^{18}$GeV, so that $g_1\sim
0.1$, we get $k_b \sim h^2/(2$kpc$)$, where we have inserted
$\eta_*\sim 20h^2$Mpc. An interesting constraint comes from the fact that the
break in the spectrum should be on a scale which is smaller than the
scale represented by the first acoustic peak in order not to reduce the
latter. Since $\eta_*$ corresponds to the horizon scale at equality,
this requires $\eta_*k_b \gtapprox 1$ or $\om_1(a_1)=H_1 \gtapprox 0.02m_{\rm
Planck}$. Together with $H_1\ltapprox 0.1 m_{\rm Planck}$, this brackets
the string scale just in the bull park where it is
expected for very different theoretical reasons.

The length-scale/energy-scale corresponding to the break $k_b$ at
the time $\eta_b$, during the pre-big bang phase, when the expansion
law is supposed to change, is given by 
\be
 |t_b| \sim |\eta_b| a(\eta_b)/a_0 \sim |\eta_b| {a(\eta_b)\over
a(\eta_1)}10^{-32} \sim
|\eta_b| \left|{\eta_b\over\eta_1}\right|^{-1/4}10^{-32}\sim 6\times
10^{-14}{\rm cm} \sim 3 {\rm GeV}^{-1}, \label{norm phys} 
\ee
where we have used $\eta_b \sim \eta_* \sim 20$Mpc and $\eta_1 \sim
0.1$cm. The energy scale obtained in this way is uncertain with a factor
of about 10.

In Fig.~\ref{break} we show the dependence of the CMB anisotropy
spectrum on the position of the break. Typically, the break lowers the
second and subsequent acoustic peaks while does not substantially
affect the first peak.

\begin{figure}[ht]
\centerline{\epsfig{file=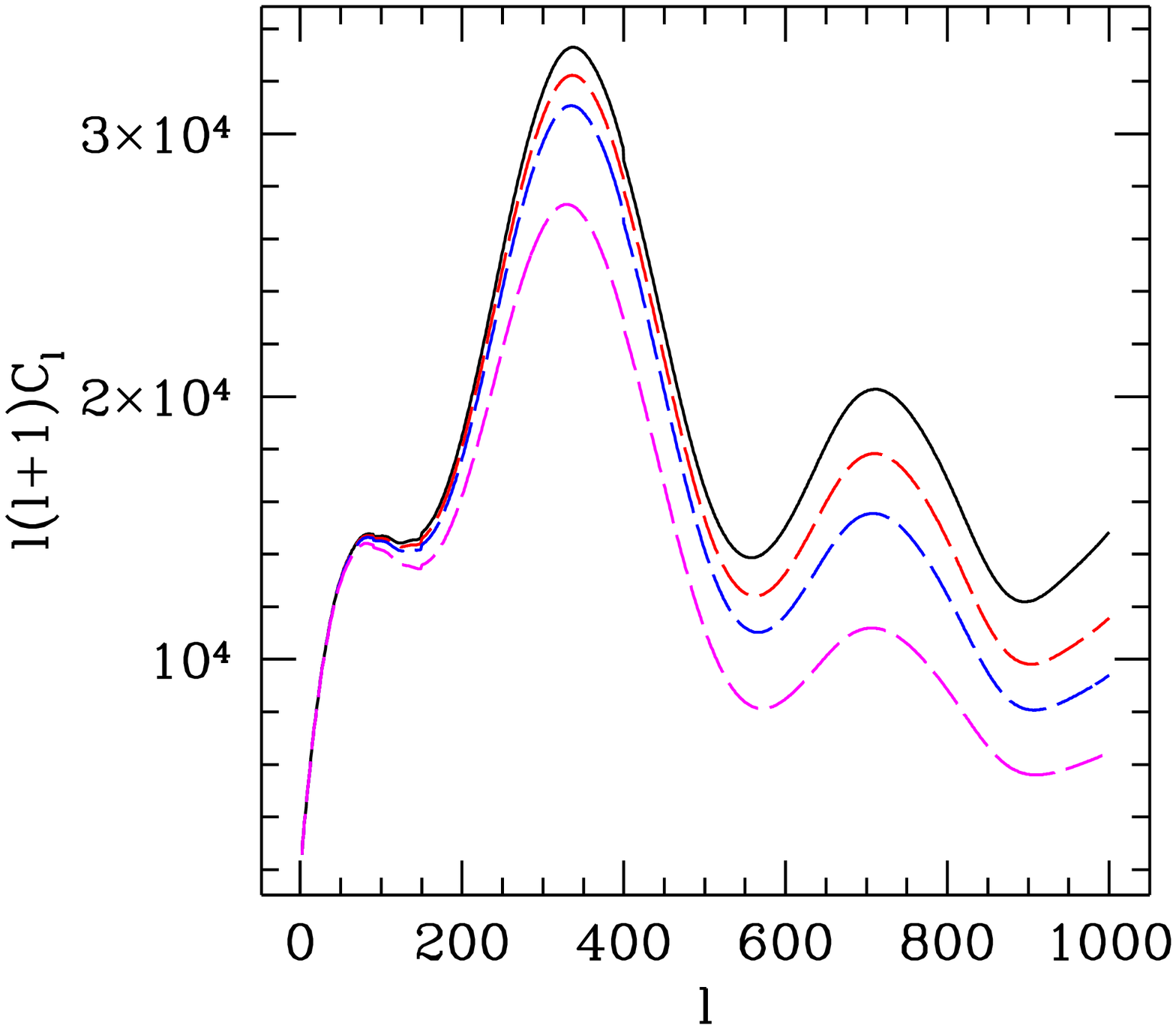, width=3.4in}}
\caption{The influence of the break position on the CMB power
spectrum. The top solid line is the spectrum without break. The dashed
lines from top to bottom represent a spectrum with break at
$k_b=3/\eta_*, 2/\eta_*$ and  $1/\eta_*$ respectively.
}
\label{break}
\end{figure}

\end{subsection}

\begin{subsection}{Cosmological parameters}

\begin{figure}[ht]
\centerline{\epsfig{file=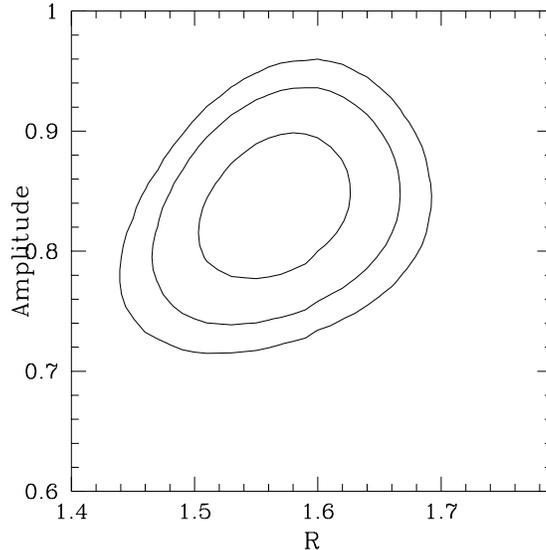, width=3in}}
\caption{Confidence levels ($68\%$, $95\%$, and $99\%$) for the 
rescaling factor $R$ and the amplitude in COBE units $A$,
from the recent BOOMERanG and MAXIMA-1 observations.}
\label{AR}
\end{figure}

\begin{figure}[ht]
\centerline{\epsfig{file=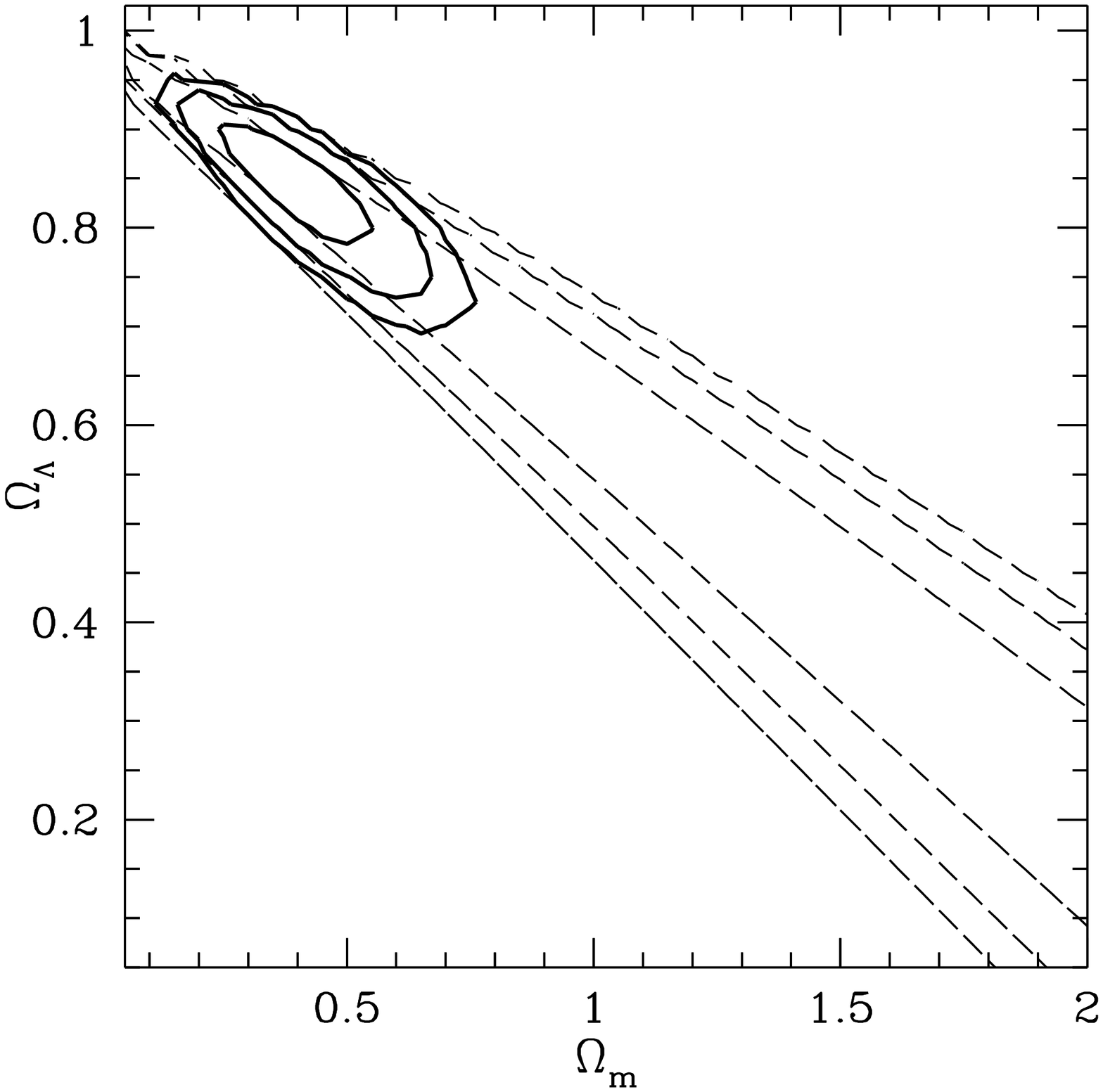, width=3in}}
\caption{The $68\%$, $95\%$, and $99\%$ confidence levels for the 
cosmological parameters
$\Omega_{\Lambda}$ and $\Omega_{m}$, from the peak position
detected by BOOMERanG and MAXIMA-1 for the
model presented in this paper (dashed). The solid contours are obtained
 including the supernovae data. \label{OmOl}}
\end{figure}

In the last two years, a peak in the CMB power spectrum at $\ell \sim 200$
as been detected by several different experiments, most notably
TOCO98~\cite{toco}, B97~\cite{b97}, B98~\cite{debbe}, and
MAXIMA-1~\cite{max}.  Among them, the BOOMERanG-98 power
spectrum~\cite{debbe} reported the best and at the same time
most conservative detection,
although coming from only $5 \%$ of their overall dataset.
The position, amplitude and shape of the peak can be fitted by
the power spectra expected in the simplest
inflationary scenario based on adiabatic perturbations in a spatially
flat universe~\cite{knpa,dlkn,melk2k,teza}. Therefore,
this peak represents the biggest challenge for the model
presented here.

We want to investigate whether a suitable choice of
cosmological parameters can bring our model in agreement with the
above mentioned data.
This question is also very important in view of
the usual ``determination of the cosmological parameters'' from CMB
anisotropies, in the sense that it shows how the results can change
when assuming a different model of structure formation.
In other words the so called ``measurements'' of cosmological parameters
from CMB anisotropies are strongly model dependent!

The peak position is determined mainly by
the angular diameter distance parameter
\be
R^{-1}=\sqrt{\frac{\Omega_m}{|\Omega_{K}|}}\frac{F(y)}{2}. 
\ee
Here $\Om_K=1-\Om_m-\Om_\La$
is the curvature parameter and 
\be
F(y)= 
\left\{
 \begin{array}{ll}
  \sinh y \ \ \ \ &{\rm (open)}  \\
   y \ \ \ \ &{\rm (flat)}   \\   
  \sin y \ \ \ \ &{\rm (closed)} 
 \end{array} \right.
\ee
depends on the geometry of the universe. The variable  $y$ is
the following integral:
\begin{equation}
y=\sqrt{|\Omega_{K}|}\int_{0}^{z_{dec}} {dz \over 
[\Omega_{m}(1+z)^3+\Omega_{K}(1+z)^2+\Omega_{\Lambda}]^{1/2}} \label{yy}.
\end{equation}

As pointed out in \cite{efbo}, the condition $R=$ constant 
identifies curves in the $\Omega_m - \Omega_\La$ plane,
 with nearly degenerate $C_\ell$ spectra, 
providing that the baryon density parameter $\Om_{\rm baryon}$ is kept constant.

In Fig.~\ref{AR} we plot likelihood contours, obtained as follows:
we rescale the string cosmology power spectra plotted in Fig.~\ref{cl}, 
both
in amplitude $A$ (in COBE units) and position $R$. We compare the
resulting spectra with the BOOMERanG and MAXIMA-1
data in the region up to $\ell \le 400$ by a simple $\chi^2$-fit.
We find that the $68 \%$ confidence limit for $R$ marginalized over
$A$ is  $ 1.50 \le R \le 1.63$ with $R=1.57$ as best fit (see Fig.~\ref{AR}).

In Fig.~\ref{OmOl} the confidence levels on $R$ are translated to
confidence levels in the $\Omega_\La - \Omega_m$ plane which are then
combined  with the current SN1a results \cite{SN1}. 
It is clear from this figure
that the model can be brought in reasonable agreement with 
observations only if the universe is closed.
The deviation from flatness becomes less and  less important
towards $\Omega_m \rightarrow 0$, where all the $R={\rm const}$ lines
converge at $\Omega_{\Lambda}=1$.
While the region with $\Omega_m>1$ can be safely excluded from
different cosmological observations, a moderately closed universe
with $\Omega_{\Lambda} \sim 0.85$ and $\Omega_m \sim 0.4$ is compatible
with SN1a results and also with estimates for $\Omega_m$ from cluster
abundance and X-ray data (see e.g.~\cite{RH}).

\begin{figure}[ht]
\centerline{\epsfig{file=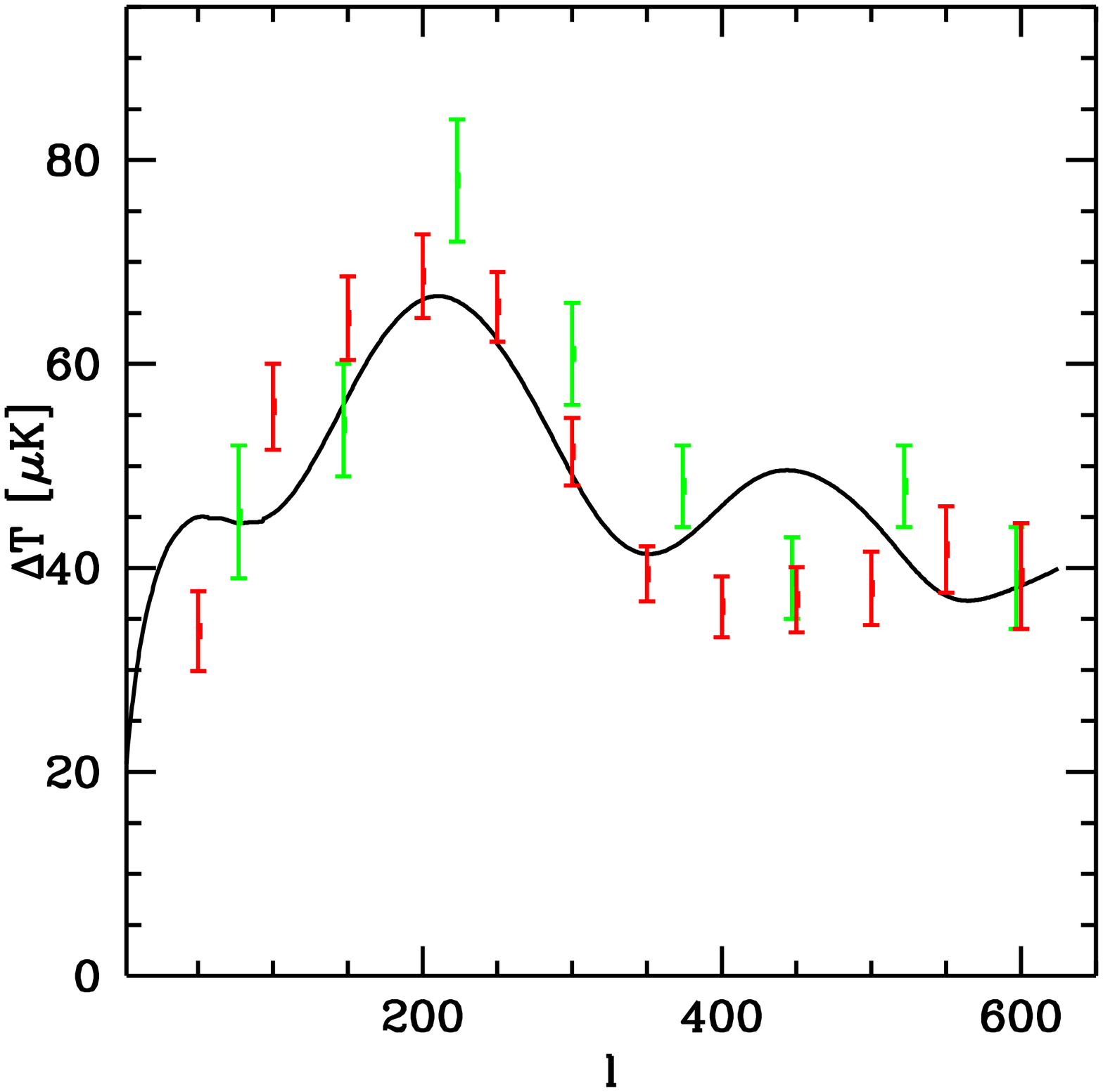, width=3in}
\epsfig{file=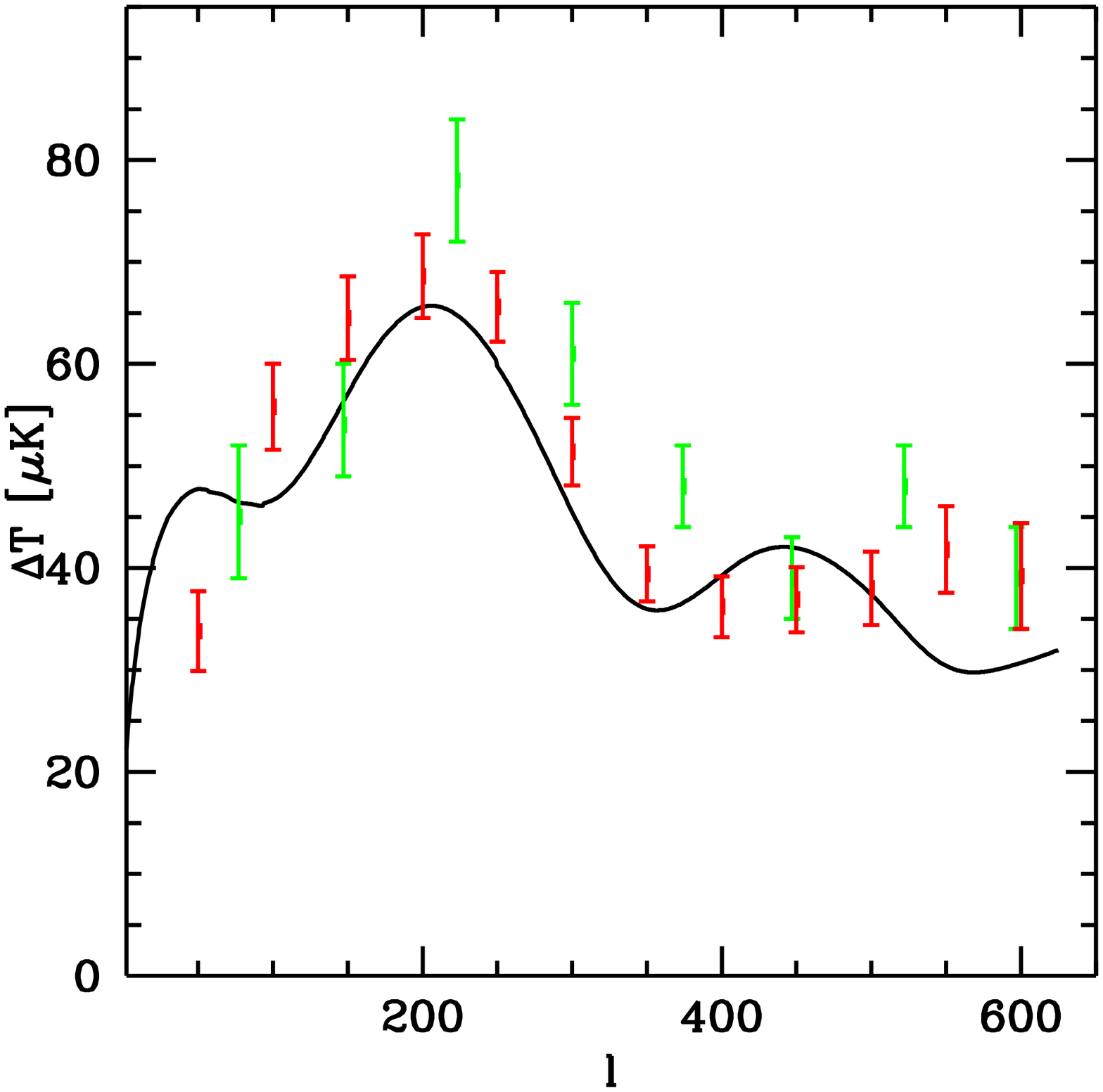, width=3in} }
%\centerline{\epsfig{file=.eps, width=3in}}
%\caption{Our best fit CMB anisotropy spectrum is compared with the
%COBE, MAXIMA and BOOMERanG98 data. This is a model with break at
%$k_b=3/\eta_*$. \label{bestfit}}
\caption{Two theoretical CMB anisotropy spectra normalized to the 
COBE data, with 
$\Omega_{\Lambda}=0.4$ and axionic spectral index $n_{\sigma}=1.33$, 
are compared with the MAXIMA-1 and BOOMERanG98 data. 
From left to right, our model 
has a  break at
$k_b=3/\eta_*$ and
$k_b=1/\eta_*$ respectively. 
Lowering $k_b$ we subtract power on small scale
and we can lower the second peak leaving the first one 
almost unchanged. \label{bestfit}}
\end{figure}

As we have seen, the position of the first acoustic peak can be adjusted
by choosing $\Omega_{\Lambda}$ and $\Omega_m$ so that the
resulting universe is marginally closed. Nonetheless, the width of
the peak, compressed by the increase of $R$, is still not in very good
agreement with the data, as well as the isocurvature hump. The 
resulting normalized $\chi^2$ is about $\sim 1.8$ for the best-fit,
which ``excludes'' the model at 70\% confidence. One has however to keep
in mind that the $C_\ell$'s are not Gaussian and therefore the
probability for our model to lead to the measured CMB anisotropies is
even somewhat higher than 30\%. In Fig.~\ref{bestfit}
two theoretical CMB spectra normalized to the COBE data 
are shown together with the MAXIMA and
BOOMERanG98 data.
%our best fit CMB spectrum is shown together with the COBE, MAXIMA and
%BOOMERanG98 data. 
%We did not optimize on the axion spectrum, the break
%position, or the baryon density parameter, but we chose $n_\si=1.33$,
%$k_b=3/\eta_*$, and $\Om_{\rm baryon}=0.05$. 
We did not optimize on the axion spectrum, 
or the baryon density parameter, but we chose $n_\si=1.33$,
$\Om_m=0.4$, and $\Om_{\rm baryon}=0.05$.

Playing with the break-scale $k_{b}$ we can in principle 
lower the second peak 
leaving the first one almost unchanged. 
Nevertheless, the position of the second peak is different from the
one indicated by inflationary models and the data. Inter-peak distance 
is therefore a better estimator of the validity of a model.
Clearly more and better
data  around the isocurvature
hump region, {\em i.e.} $\ell\sim 100$, is needed to decide definitely
whether the model is ruled out. This will most probably be achieved
with the MAP satellite~\cite{Map} planned for launch in 2001. 
\end{subsection}

\begin{subsection}{Polarization}

\begin{figure}[ht]
\centerline{\epsfig{file=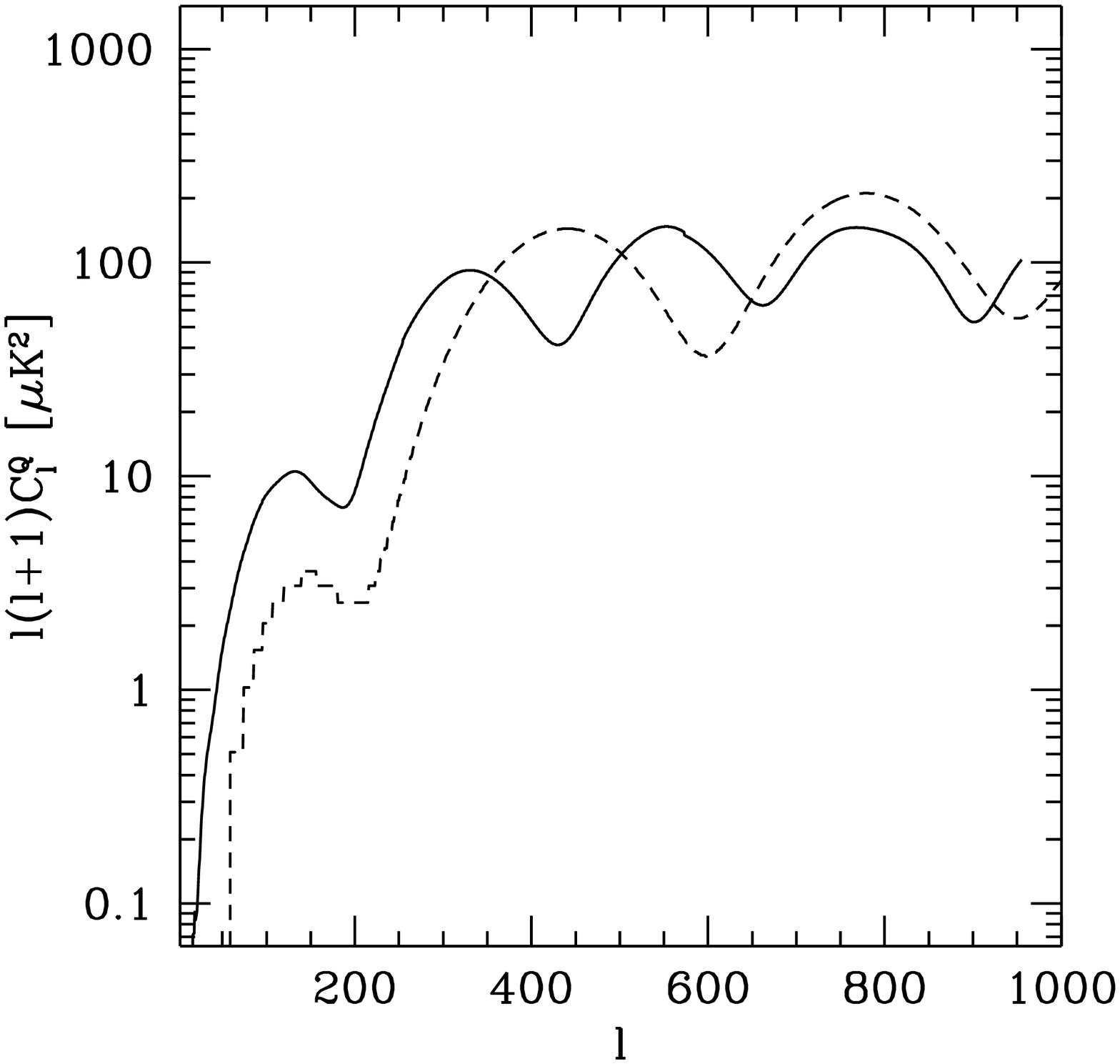, width=3in}}
\caption{The CMB polarization spectrum of our model (solid line) for
the best fit parameters is compared with the inflationary  CMB
polarization spectrum  in a critical universe with $\Om_\La=0.7$. The fact
that in our model the universe is closed is visible in the smaller
distances between successive peaks.  \label{bestfitpol}}
\end{figure}

The polarization spectrum distinguishes easily between 
adiabatic inflation and the axion seed model (see Fig.\ \ref{bestfitpol}). 
The preferred closed universe
for axion seeds translates into a smaller distance between
polarization peaks. As the physical distance between peaks depends
only on the sound speed, which is only slightly dependent on $\Om_{\rm
baryon}h^2$, a quantity which is already tightly constrained by
nucleosynthesis, the $\De \ell$ on which this distance projects is
mainly determined by spatial curvature, $\Om_K$ (it
depends also somewhat on
$\Om_\La$ as can be seen from Eq.~(\ref{yy})), and is independent on
the model for the initial fluctuations. 

\end{subsection}

\begin{subsection}{The dark matter power spectrum}
\begin{figure}[ht]
\centerline{\epsfig{file=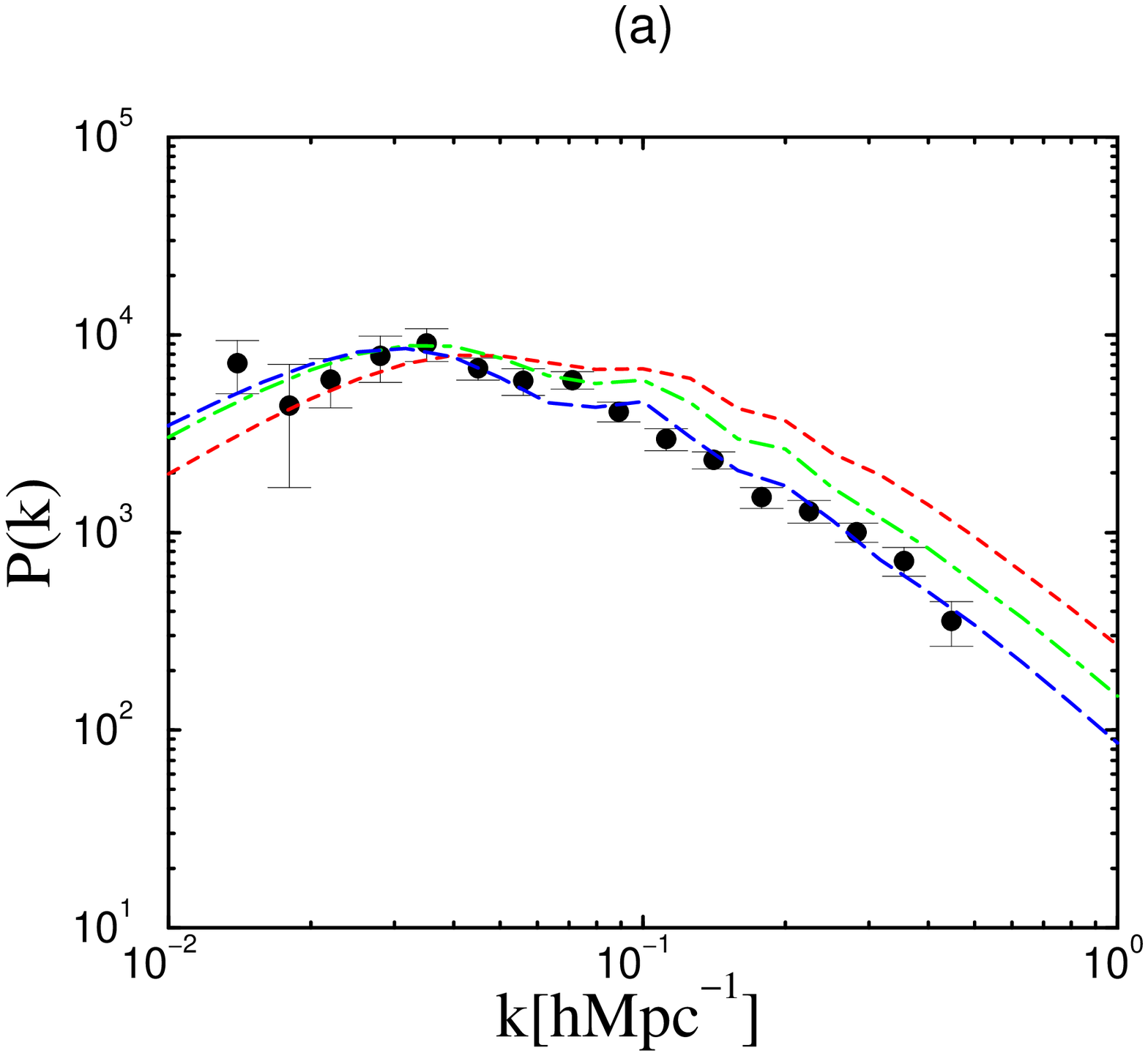, width=3.3in} 
\epsfig{file=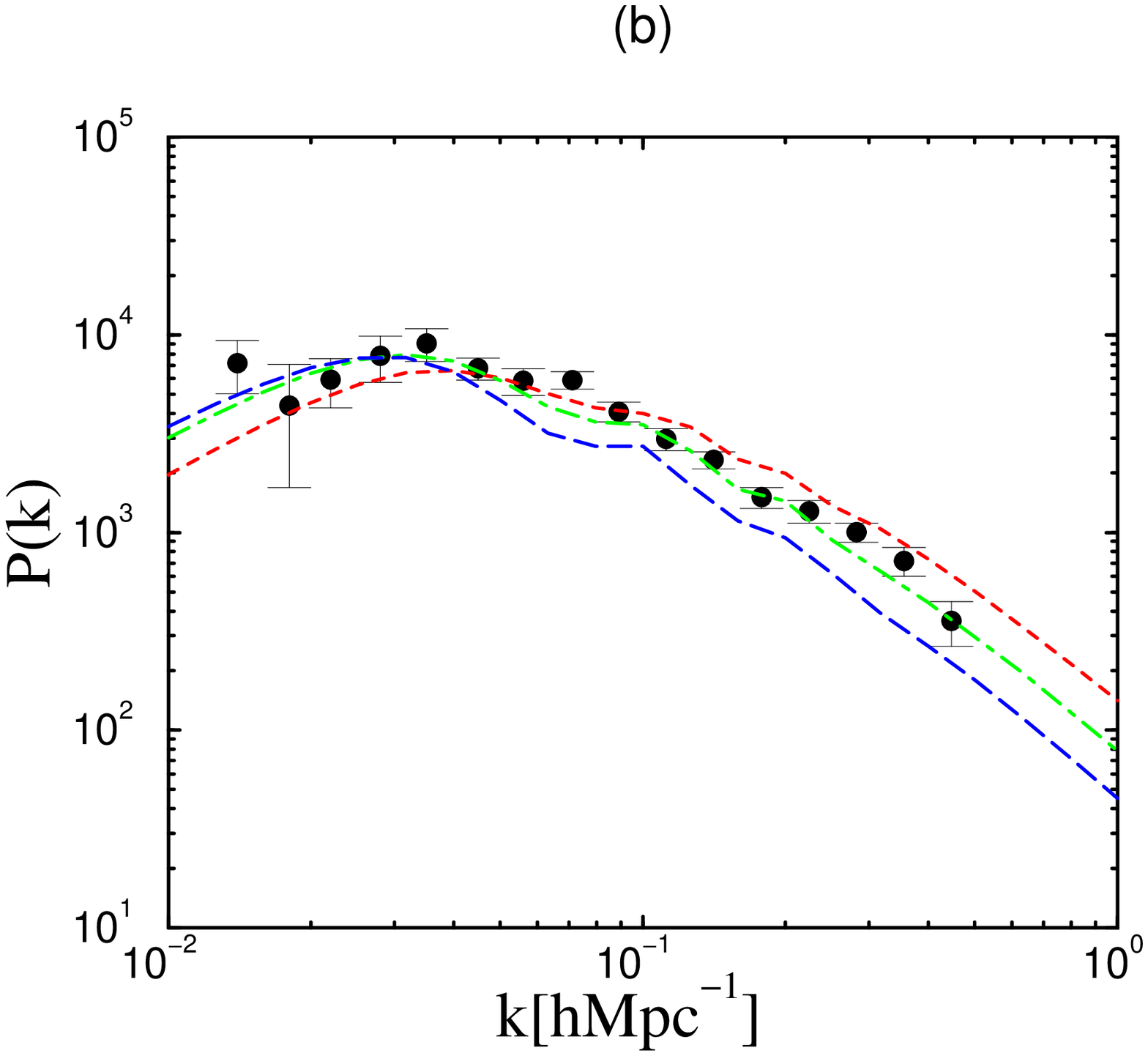, width=3.3in}   }
\caption{The linear dark matter power spectra for fluctuations induced
by axion seeds with spectral index $n_{\sigma}=1.33$ and a break 
in the spectrum at (a) $k_b=3/\eta_*$ and (b) $k_b=1/\eta_*$, 
for a flat universe
with $\Omega_m=0.4$ (dotted), $\Omega_m=0.3$ (dot dashed), and
$\Omega_m=0.25$ (dashed) are compared with data. We assume an 
IRAS galaxies bias of $b_I=\Omega_m^{-0.3}$.}
\label{DM}
\end{figure}

The computation of the dark matter power spectrum had already been performed 
in \cite{afrg} where a considerable deviation 
from the data was found. In this work we repeat this computation 
taking into account the preferred values of the axion spectral index 
and of the matter energy density found from CMB data,
and we introduce the break in the axion spectrum discussed above. 
With this additional input it is possible to
establish reasonable agreement between the data and the dark matter
power spectrum (see Fig.~\ref{DM}). 

Since the computation of the theoretical 
matter power spectrum for a closed universe
is relatively involved and since, for the purpose of comparing the
theoretical spectrum with observations, we are interested in scales
much below the curvature scale, we have computed it 
for a flat universe, with matter and a cosmological constant,  
assuming that the contribution from curvature is negligible on the
scales under consideration. Indeed, what 
really plays a role for the matter power spectrum
is the matter content, $\Omega_m$, which fixes the 
time of equality between matter and radiation, determines when
structures can start growing, and fixes the position of the bend in the
power spectrum.

In Fig.~\ref{DM} we present the theoretical 
dark matter power spectra together with the data as compiled by 
Peacock and Dodds \cite{Pea}. Depending on the scale of the break
in the axion spectrum, 
$\lambda_b =1/k_b$, our model can be compatible with data for different 
values of $\Omega_m$  in the range
$0.2\leq \Omega_m \leq 0.4$.
The role of the break is the following: if $\lambda_b$ is small we
subtract power only from small scales and we are able to reproduce a
power spectrum in good agreement with data even if $\Omega_m$ is
relatively high. However, if we do not introduce any  break in the
axion spectrum we find too much power on small scales and 
our theoretical dark matter power spectrum is incompatible with data  
(compare our present result with those 
found in \cite{afrg}).
 
The root mean square mass fluctuation within a ball of radius
$8h^{-1}$ Mpc for the model with $n_{\sigma}=1.33$, $k_b=3/\eta_*$, and 
$\Omega_m=0.25$ and for the model with $n_{\sigma}=1.33$, $k_b=1/\eta_*$, 
and 
$\Omega_m=0.4$ are $\sigma_8 = 0.85$ 
and $\sigma_8 = 0.74$ respectively. 
Analysis of the abundance of galaxy clusters suggests 
$\sigma_8 \sim 0.5 \Omega_{m}^{-0.5}$ 
\cite{Eke}.
\end{subsection}

\begin{subsection}{Conclusions}
We have shown that it is possible to choose  cosmological
parameters which bring our model in reasonable agreement with the
present CMB anisotropy measurements, which is however less favorable than
the striking fit of simple flat adiabatic inflationary models.
This is our main result.

Even if our model will turn out to disagree with better data, we
believe that we learn the important lesson that cosmological
parameters obtained from  CMB anisotropies are strongly model
dependent, a point which is  swept under the carpet by the vast
majority of the circulating ``parameter-fitting'' literature. 
We believe that it is very important in the future to concentrate on model
independent quantities, like inter-peak distances, to determine
cosmological parameters.
\end{subsection}

\end{section}
\vspace{1cm}

\begin{section}{Gravitational waves}
Gravitational waves represent one of the most powerful tools to 
investigate the early history of the universe.
They decouple at a temperature comparable to the string scale which
makes them an important window for cosmological phenomena related to
the string theory domain. 
In this section we show that axions can contribute substantially 
to the production of the gravitational wave background in the pre-big
bang model, acting as a source in the tensor perturbation equation.
This leads
to a spectrum which is different from the standard gravitational wave
background of string cosmology based
on the ``direct mechanism'' of graviton production by    
amplification of quantum vacuum fluctuation. 
This new ``indirect mechanism'' leads to a flat spectrum
and can easily be distinguished from the direct one.
Indeed, as we shall see, the axion induced gravity wave background dominates
the ``direct background'' at small frequencies and 
represents an
important observational constraint for string cosmology.

\begin{subsection}{Direct production -- Amplification of vacuum fluctuations}

So far, amplification of quantum 
vacuum fluctuations have been considered as the 
principal mechanism for the production of gravitational waves during the 
pre-big bang phase \cite{Gaspgiov,Gaspven,Brumbrum}.
During the dilaton era, before the big bang,
when the scale factor evolves according to \Eq{Solu}, 
the Fourier modes of metric tensor perturbations
satisfy an evolution equation similar to \Eq{Evol}, namely
\be
\ddot{\psi}^T_{\bk} +
\left(k^2-{\ddot{a}_T\over a_{T}}\right)\psi^T_{\bk} =0, 
\label{Evolgrav}
\ee 
where $a_T=ae^{-\phi /2}$ is the pump field of gravity waves
and $\psi^T_{\bk}$ is the canonical
variable for tensor modes of the metric.
For the isotropic case discussed in this work, one find that 
$a_T \propto |\eta|^{1/2}$ independently on the evolution and 
number of dimensions during the pre-big bang phase.
After proper normalization to the incoming vacuum, this yields the solution
\be
\psi^T_{\bk}=(-\eta)^{1/2} H^{(2)}_{0}(k \eta), \ \ \ \ \eta < -\eta_1.
\ee

After the big bang, in the radiation dominated era, $\eta > \eta_1$,  
the solutions 
of  \Eq{Evolgrav} are simple plane waves. 
From the matching conditions 
between these two regimes, applying the same procedure as discussed
in Subsection~\ref{Ampli} for the axion field, 
one obtains 
the following spectrum of gravitational waves, 
\be
\Omega_g \sim \frac{\omega_1^4}{H_0^2m_{\rm Planck}^2} 
\left(\frac{\omega}{\omega_1}\right)^3 
\sim g_1^2 
\left(\frac{\omega}{\omega_1}\right)^3 \Omega_{\gamma},
\ee
which is a tilted spectrum, $\propto \omega^3$, normalized to $g_1^2$
at the string scale. One 
actually supposes that, at a string epoch $\eta_s < -\eta_1$, 
the dilaton-vacuum 
regime behavior of \Eq{Solu} breaks down and the universe undergoes 
a De Sitter 
expansion with linearly growing dilaton, which
lasts until the beginning of the radiation dominated era $\eta_1$. 
This phase leads to a nearly flat
gravitational wave spectrum at very small scales. The normalization
of the spectrum to the string coupling $g_1$ can then
be performed at a lower frequency, $\omega_s < \omega_1$, 
leading to a somewhat higher density of directly produced gravitons
than the one discussed here. This is very important in order
to make the direct
background observable and still compatible with nucleosynthesis. 
(See \cite{Michele} and references therein for more details.)
A more detailed discussion on the important signatures
and observational consequences of this direct production of
gravitational waves can be found in \cite{Gaspero} and references therein. 

\end{subsection}
\begin{subsection}{Indirect production -- Axion source}

Let us discuss now the production 
of a stochastic gravitational wave background generated 
by the presence of axion seeds.
This indirect background will be superimposed to the direct one discussed 
above and will dominate the total spectrum at large scales.
These two production mechanisms are fundamentally different.
While the direct production of gravitons
takes place during the pre-big bang phase and is due to the amplification
of vacuum fluctuations, the indirect production is sourced by the axions
during the post-big bang era. 

The creation, propagation, and damping of gravitational waves 
in a Friedman background are described by the tensor perturbation equation 
(see e.g. \cite{MTW}),
\be
\ddot{h}_{ij}+2\frac{\dot{a}}{a} \dot h_{ij} -\Delta
h_{ij} =16 \pi G a^2\tau_{ij}, 
\label{Tens1}
\ee
where tensor perturbations in
the metric are parameterized by the traceless, divergence-free, symmetric
tensor field $h_{ij}$,
\be
g_{\mu \nu}=\overline{g}_{\mu \nu}+ a^2(\eta) h_{\mu \nu}, \ \ \ \ \ \  
h^{\mu}_{\mu}=0=\nabla_{\nu}h^{\nu}_{\mu},
\ee
which is a gauge invariant variable. As 
before a dot denotes the derivative with respect to
conformal time.
\Eq{Tens1} is a wave equation with source term $\tau_{ij}$.

The tensor field $h_{ij}$ is usually decomposed into two 
polarization states as
\be
h_{ij}(\bx,\eta)=h^{\times}(\bx,\eta)\epsilon^{\times}_{ij}(\bx) 
+ h^{+}(\bx,\eta)\epsilon^{+}_{ij}(\bx),
\ee
where $\epsilon^{\times}_{ij}=
\bbe^1_i\bbe^1_j-\bbe^2_i\bbe^2_j$ and $\epsilon^{+}_{ij}=
\bbe^1_i\bbe^2_j+\bbe^2_i\bbe^1_j$ are the polarization tensor fields and
$(\bbe^1,\bbe^2,\bbe^3)$ is a local orthonormal basis 
(the wave is propagating in the $\bbe_3$ direction). 

The energy density of gravitational waves is given by the 
00-component of the energy momentum tensor of the wave. This can be defined as
a space-average over several oscillations,
\be
\rho_g=\frac{\langle \dot{h}_{ij} \dot{h}^{ij} \rangle}{16 \pi Ga^2} =
\frac{\langle \dot{h}_{\times}^2\rangle + 
\langle \dot{h}_{+}^2 \rangle}{16 \pi Ga^2}.
\ee
We decompose $h_{\times}$ and $h_{+}$ in Fourier 
modes,
\be
h_{\lambda}(\bx,\eta)=\int \frac{d^3k}{(2 \pi)^3} 
e^{i\bk\cdot\bx} h_{\lambda}(\bk,\eta), \ \ \ \ \lambda=\times,+; 
\ee
therefore 
\be
\dot{h}_{\lambda}(\bx,\eta)= \int \frac{d^3k}{(2 \pi)^3} 
e^{i\bk\cdot\bx} \dot h_{\lambda}(\bk,\eta).
\ee
The spatial average then becomes
\be
\langle \dot{h}_{\lambda}^2 \rangle=
 \int \frac{d^3k}{(2 \pi)^3} \frac{d^3k'}{(2 \pi)^3}
e^{i \bx \cdot (\bk + \bkp)} 
\langle\dot h_{\lambda} (\bk,\eta)\dot h_{\lambda} (\bkp,\eta) \rangle,
\ee
and we can use the stochastic average condition
\be
\langle \dot{h}_{\lambda} (\bk) \dot{h}_{\lambda'} (\bkp) \rangle 
= (2\pi)^3 \delta^3(\bk - \bkp)\de_{\la\la'} |\dot h_{\lambda}(\bk)|^2,
\ee
which yields, under the hypothesis of statistical isotropy,
\be
\rho_g=\frac{1}{(\pi a)^2 16\pi G}\int dk k^2 |\dot h_{\lambda}(k,\eta)|^2.
\label{Grav1}
\ee

We now compute the spectrum 
$|\dot h_{\lambda}(k,\eta)|^2$ in the coherent approximation. For this we
introduce the deterministic source function $\Pi(k,\eta)$ defined by
\be
4 \pi G a^2\Pi(k,\eta) \equiv \sqrt{H(k,\eta,\eta)},
\ee
(for the function $H$, see \Eq{Tens12}).
The polarization tensors satisfy 
$\epsilon^{\lambda}_{ij} \epsilon^{ij}_{\lambda'}=
2\delta_{\lambda}^{\lambda'}$ and
we can hence rewrite \Eq{Tens1} in momentum space as
\be
\ddot{h}_{\lambda}+2\frac{\dot{a}}{a}\dot{h}_{\lambda}+k^2 h_{\lambda}
=8 \pi G a^2 \Pi.
\label{Tens3}
\ee
The factor $1/2$ comes from the fact that $\Pi$ sources both modes $\times$ 
and $+$ of
$h_{ij}$ and, assuming again statistical isotropy, each mode is sourced with
the same strength.
Since we want to compute a gravitational wave spectrum 
we only consider modes which enter the horizon in the
radiation dominated era, $k\eta_{*} > 1$ and $\dot{a}/a \simeq 1/\eta$,
the other modes being uninteresting (too large wavelength) for
possible observations.
Therefore we also consider modes far from COBE scale, 
$k\gg k_b$, and we can comfortably assume a 
flat axion spectral index, $n_{\sigma}=1$.
We then write \Eq{Tens3} as
\be
 h''_{\lambda}+\frac{2}{x}h'_{\lambda}+h_{\lambda}=\left\{
\begin{array}{ll}
 f(k)/\sqrt{x} \ \ \ &x\le 1 \ \ \  
\mbox{(active source)}\\
0 \ \ \ \ \ \ \ \ \ \ \ &x\ge1\ \ \
\mbox{(dead source),}
 \end{array} \right.
\ee
where the conformal time derivative has been replaced by the derivative with respect 
to $x=k\eta$. 
In this equation we assume that the axion source can 
be approximated by a power law behavior outside the horizon which is
of the form
\be
8\pi G a^2 \Pi(k,x)=x^{-1/2}k^2f(k), \ \ \ \ \ f(k)\simeq 8 \pi g_1^2 k^{-3/2},
\ee
and can be considered negligible inside the horizon where the correlators 
decay quickly.

The homogeneous solutions to this equation are the spherical Bessel functions
 of index zero, $j_0(x)$ and $y_0(x)$.
In the regime, $x\le1$, the solutions can be found with
the Wronskian method, which yields
\be
h_{\lambda}(k,x)=f(k) [c_1(x) j_0(x) +c_2(x) y_0(x)], \ \ \ \ x \le 1,
\label{hsup}
\ee
where
\be
c_1(x)=\int_0^1 dx x^{1/2}\cos x, \ \ \ \ c_2(x)=\int_0^1 dx x^{1/2}\sin x,
\ee
while in the second regime, $x\ge1$, they are a linear
combination of the homogeneous solutions, 
\be
h_{\lambda}(k,x)=A(k)j_0(x)+B(k)y_0(x) \ \ \ \ x \ge 1.
\label{hsub}
\ee
By matching Eqs.~(\ref{hsup}) and (\ref{hsub}) at $x=1$ we find
\be
h_{\lambda}(k,\eta)=f(k)[c_1(1)j_0(k\eta)+c_2(1)y_0(k\eta)], 
\ \ \ A(k)=f(k)c_1(1), \ \ \ B(k)=f(k)c_2(1)
\ee
which yields, for $x \gg 1$, $\dot h_\la \sim kh_\la\sim
   f(k)/\eta$, and thus
\be
|\dot h_{\lambda}(k,\eta)|^2\simeq \frac{(8 \pi)^2 g_1^4}{\eta^2}k^{-3}.
\ee
Using \Eq{Grav1} we hence find 
\be
\rho_g = \frac{4g_1^4}{\pi G a^2 \eta^2}\int\frac{dk}{k}, \ \ \mbox{or} 
\ \ \ \ \ \ \frac{d\rho_g}{d\log k}=\frac{4g_1^4}{\pi G a^2 \eta^2},
\ee
which corresponds to a flat spectrum of gravitational waves.

On the other hand, at early time the radiation energy density, 
$\rho_{\gamma}$, dominates the 
Friedman equation which becomes
\be
\frac{\dot{a}^2}{a^2}=\frac{8 \pi G}{3} \rho_{\gamma}a^2.
\ee
With $\dot{a}/a\simeq1/\eta$ we can write the gravitational wave background
spectrum produced by the axion field as
\be
\Omega_g = \frac{\rho_g}{\rho_{\gamma}} \Omega_{\gamma} 
\sim 10 g_1^4 \Omega_{\gamma}.
\ee

\begin{figure}[ht]
\centerline{\epsfig{file=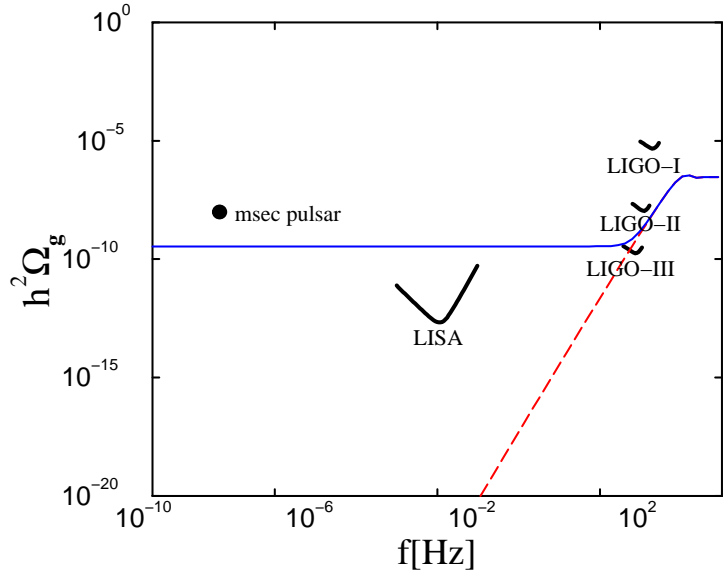, width=4in}}
\caption{The gravitational wave spectrum in pre-big bang model for 
a value of $g_1=0.03$. 
The directly produced background (dashed line) 
has been
normalized to string mass at a frequency of $f_s=500{\rm Hz}$. This 
frequency corresponds to the time $\eta_s = - 1/(2\pi f_s a(\eta_0))$  
for the transition between the dilaton-dominated regime and 
the De Sitter phase in the pre-big bang era (see above). 
The solid line represents the sum of the direct and indirect production of 
gravitational wave background. The analysis has been limited to 
frequencies $f\gg f_b$, where $f_b$ denotes the frequency
corresponding to the scale of the break.}
\label{gravit}
\end{figure}
\end{subsection}

\begin{subsection}{Observational consequences}

In the previous subsection we derived the spectrum of gravitational waves 
induced by axion seeds and we found that 
it is flat on scales much smaller
than the COBE scale and normalized such as to lead to the correct 
amplitude of fluctuations in the CMB anisotropies.

Its normalization 
depends on the fundamental ratio between the string and Planck mass which 
is usually taken to be of the order of $g_1 \sim 0.1 \div 0.01$ \cite{g1}.
The energy density of induced gravity waves is proportional to the
forth power of $g_1$ like the CMB anisotropy spectrum.
Since the COBE normalization also depends on $k_b$ (see \Eq{norm}), which plays
no r\^ole for the gravity wave spectrum on the scales considered here,
$g_1$ alone is still allowed to vary in the range cited above even
though \Eq{norm} provides a precise constraint for a combination of $g_1$
and $k_b$. Using the previous values for $g_1$ we find a flat spectrum
of gravity waves with  $h^2\Omega_g \sim 4\times (10^{-8} \div
10^{-12})$, a range which, most probably, will be reached by  
the third generation interferometers~\cite{mm}. This renders the 
indirect gravity wave background an important observable of string
cosmology. Note also that in the case of its detection it would provide
a direct measurement of the string scale!

At present
the most relevant observational bound for a gravity wave background
comes from  pulsars. In particular, the timing of the milli-second
binary pulsar implies a limit on any stochastic gravity wave background
of $h^2\Omega_g({\rm at} f=4.4\times 10^{-9}{\rm Hz})<1\times 10^{-8}$ 
(at 95 \% c.l.) \cite{Pulsar}, 
which transforms in our case 
into a limit on $g_1 \ltapprox 0.07$ in this model.

The direct gravitational wave background 
has a blue spectrum and therefore dominates the indirect background 
on small scales,  as shown in 
Fig.~\ref{gravit}\footnote{Sensitivity curves for 
LISA and LIGO are based on \cite{mm,cu} and references therein. 
We acknowledge Carlo Ungarelli.}.
The  crossover frequency $\omega_c$ between the two regimes is
determined by $g_1$ and
the normalization frequency $\omega_s$ discussed above,
\be
\omega_c = g_1^{2/3} \omega_s.
\ee
This crossover may actually, depending on the unknown value $\omega_s$, 
fall into the range of frequencies at which interferometers
will be operating.

Finally, we would like to point out that, like the CMB anisotropies of
this model, the indirect gravity wave background 
considered here is not Gaussian, which can lead to interesting 
observational consequences. 
\end{subsection}

\end{section}
\vspace{1cm}

\begin{section}{Conclusions}
We have investigated the consequences of  axion seeds which naturally
occur in  the context of string
cosmology. We found that these seeds may induce the observed large
scale structure and CMB anisotropies in the universe provided that
there is a break in the primordial axion power spectrum which from 
slightly blue on very large scales  turns to a flat spectrum
on scales smaller than the break, $k>k_b$. Such a break
appears if the expansion law undergoes a transition during the pre-big
bang phase. For the scenario to agree with observations the break must
occur at $\eta_b \sim -0.3\eta_*$, which corresponds to an
energy scale of the order of several GeV. 

The axion seed model leads to isocurvature fluctuations with important
contributions from vectors (about 50\%) and tensors (about 15\%) on
large scales. The first acoustic peak in the CMB anisotropy 
power spectrum is 
around $\ell \sim 300$ for a
flat model, $\Om=\Om_\la+\Om_m=1$. To reproduce observations the
universe has to be closed with parameters, $\Om_\La \sim 0.85$ and
$\Om_m\sim 0.4$. This parameter choice is also in agreement with
supernovae and cluster data. Even though our model leads to a larger
$\chi^2$ when fit to the CMB data it cannot be excluded by the
presently available data. However, the ``isocurvature hump'' at $\ell\sim
40$ and the reduction not only of the second but also of the third
acoustic peaks are signatures which clearly distinguish the model from
standard inflationary scenarios. Furthermore the CMB polarization spectrum
significantly differs from the inflationary result.

We have also studied gravitational waves which are generated during the
post-big bang phase by the tensor type anisotropic stresses in the
energy-momentum tensor of the axion field. We found that they lead to
a flat observable background of gravity waves which can give
stringent constraints on the model if detected by the planned LIGO-III 
and LISA observatories. 

As the model studied is very predictive let us finally mention that
its failure to reproduce observational data, which is hinted by present
CMB anisotropy measurements and might be reinforced by future more
accurate data, does not by itself rule out string cosmology. An
additional important hypothesis of the model is that non-gravitational
interactions of the axion field with the dark matter may be neglected
and the axion plays the role of a ``seed''. If this hypothesis is
relaxed, the axions may interact with radiation and dark matter
and even lead to a standard adiabatic fluctuation
spectrum. This idea deserves further study, but most probably the
non-Gaussian character of the perturbations also survives in such a scenario.  

\begin{subsection}*{Acknowledgments}
We are grateful to Gabriele Veneziano for stimulating discussions. We 
acknowledge Cyril Cartier, Maurizio Gasperini, Martin Kunz and
Carlo Ungarelli for helpful comments. 
This work has been supported by the Swiss NSF.
One of us (F.V.) 
acknowledged financial support from the Universit\`a di Padova.
\end{subsection}
\end{section}
%%%%%%%%%%%%%%%%%%%%%%%%%%%%%%%%%%%%%%%%%%%%%%%%%%%%%%%%%%%%%%%

\end{document}